\documentclass[entropy,article,accept,moreauthors,12pt,a4paper]{mdpi} 
\lastpage{4955}
\doinum{10.3390/e15114932}
\pubvolume{15}
\pubyear{2013}
\history{Received: 24 September 2013; in revised form: 30 October 2013 / Accepted: 31 October 2013 / \\
Published: 13 November 2013}

\usepackage{subfigure}
\makeatletter
\renewcommand{\@thesubfigure}{\normalsize(\textbf{\alph{subfigure}})}
\makeatother

\usepackage{soul,booktabs}

\usepackage{enumerate}
\usepackage{amsmath}

\usepackage{amssymb}
\usepackage{rotating}
\usepackage{pdflscape}
\graphicspath{{./Figures/}}
\newenvironment{changemargin}[2]{%
\begin{list}{}{%
\setlength{\topsep}{0pt}%
\setlength{\leftmargin}{#1}%
\setlength{\rightmargin}{#2}%
\setlength{\listparindent}{\parindent}%
\setlength{\itemindent}{\parindent}%
\setlength{\parsep}{\parskip}%
}%
\item[]}{\end{list}}

\Title{Dynamical Structure of a Traditional Amazonian Social~Network}

\Author{Paul L. Hooper $^{1,2,}$*, Simon DeDeo $^{1,3,}$*, Ann E. Caldwell Hooper $^{2}$, Michael Gurven $^{4}$ and \mbox{Hillard S. Kaplan $^{5}$}}

\address{%
$^{1}$ Santa Fe Institute, 1399 Hyde Park Road, Santa Fe, NM 87501, USA\\
$^{2}$ Department of Anthropology, Emory University, 1557 Dickey Drive, Atlanta, GA 30322, USA; \linebreak {E-Mail: anncwell@gmail.com}\\
$^{3}$ School of Informatics and Computing, Indiana University, 901 E 10th Street, Bloomington, \linebreak IN 47408, USA \\
$^{4}$ Department of Anthropology, University of California Santa Barbara, Santa Barbara, CA 93106, USA; {E-Mail: gurven@anth.ucsb.edu} \\
$^{5}$ Department of Anthropology, University of New Mexico, MSC01-1040, Albuquerque, NM 87131, USA; {E-Mail: hkaplan@unm.edu}
}

\corres[2]{E-Mails: phooper@santafe.edu (P.L.H.); simon@santafe.edu (S.D.); {Tel.: +1-505-984-8800; Fax: +1-505-982-0565.}}

\abstract{Reciprocity is a vital feature of social networks, but relatively little is known about its temporal structure or the mechanisms underlying its persistence in real world behavior. \mbox{In pursuit of} these two questions, we study the stationary and dynamical signals of reciprocity in a network of manioc beer (Spanish: \emph{chicha}; Tsimane': \emph{shocdye'}) drinking events in a Tsimane' village in lowland Bolivia. At the stationary level, our analysis reveals that social exchange within the community is heterogeneously patterned according to kinship and spatial proximity. A positive relationship between the frequencies at which two families host each other, controlling for kinship and proximity, provides evidence for stationary reciprocity. Our analysis of the dynamical structure of this network presents a novel method for the study of conditional, or non-stationary, reciprocity effects. We find evidence that short-timescale reciprocity (within three days) is present among non- and distant-kin pairs; conversely, we find that levels of cooperation among close kin can be accounted for on the stationary hypothesis alone.\\}

\keyword{social systems; social networks; time-series analysis; cooperation; reciprocity; kinship; Tsimane'; forager-horticulturalists}

\begin{document}

\begin{changemargin}{1cm}{1cm}

\small
\noindent \emph{Se hac\'{i}a chicha, despu\'{e}s se juntaba la gente y despu\'{e}s todita la gente estaba borracha y cantaban las canciones de los animales: del chancho de tropa, del marimono, del maneche, y segu\'{i}an cantando...} 

\noindent \emph{Chicha} was made, and the people got together, and then everyone was drunk and singing the songs of the animals---the white-lipped peccary, the spider monkey, the howler monkey---and on they sang... \\

\noindent \hfill --Felipe Huallata \cite{iamele} 
\end{changemargin}

\section{Introduction}

The evolutionary social and life sciences seek to understand both the ultimate evolutionary logic that motivates the formation of valuable social relationships, as well as the proximate mechanisms that support the stability of these relationships over time \cite{silk2012,jackson2002,kappeler2013}. Longitudinal social network data provide an arena for evaluating the predictions of evolutionary models for relationship formation and maintenance through time. These data also inform us of the time dimension of network activity, a topic of both empirical and theoretical interest in the network sciences \cite{doreian1997,newman2009}. More generally, these data can shed light on the time scale of the processes that are in motion within social and biological systems \cite{flack2012,dedeo1,dedeo2}.

Among the indigenous farming populations of the Amazon, social gatherings centered around drinking home-brewed manioc beer (Spanish: \emph{chicha}; Tsimane': \emph{shocdye'}) provide a context for the development and exercise of valuable cooperative relationships \cite{mowat}. In addition to the food value of the beer, conviviality and intoxication create a social venue for conversation, exchange and friendship. Investment in these social bonds may yield benefits in other domains of social life, such as cooperative labor, food sharing, politics, mobility and mate search. 

Here, we analyze time-series data on hosting and attending manioc beer events collected across a sixteen-week period in a Tsimane' village in Bolivian Amazonia. We use multi-level modeling methods to analyze the stationary patterns of social events across the sample period. We evaluate the extent to which these patterns are consistent with predictions from the evolutionary theories of kin selection---that individuals behave more altruistically toward closer kin \cite{hamilton1964}---and reciprocity---that favors given are repaid by favors received in return \cite{trivers1971}. For our analysis of the nature of the underlying dynamical processes that create and sustain these stationary patterns, we formulate an account of reciprocity in terms of conditional probabilities and test for statistical significance using a hierarchy of null models. These models evaluate the extent to which family $i$ hosting family $j$ is predictive of $j$ hosting $i$ in return within a specified period of time.

\subsection{Evolutionary Background and Motivation}

Evolutionary theory suggests a number of pathways by which selection can be expected to favor the emergence and stability of altruistic and cooperative social relationships \cite{gurven2004,west2007}. Hamilton's theory of kin selection \cite{hamilton1964} (which predicts that altruism is favored when the cost to donors is less than the benefit to recipients devalued by the coefficient of relatedness, $r$) provides a fundamental anchor-point for understanding the stability of many observed cases of altruism in the natural \mbox{world \cite{langergraber2012,griffin2003,bourke2011}.} Anthropologists have similarly recognized the central organizing role of kinship in structuring patterns of social interaction in human societies \cite{service1962}; and a growing number of quantitative studies have demonstrated the explanatory power of relatedness in traditional field settings \cite{hames1987,gurven2004,hooper2011,koster2011interhousehold}. 

The theory of dyadic reciprocity, articulated in sociology by Mauss \cite{mauss} and developed in evolutionary terms by Trivers \cite{trivers1971} and Axelrod and Hamilton \cite{axelrod1981}, provides an additional pathway to cooperation, which may operate independently of or in interaction with genetic kinship. The stability of cooperation through reciprocity is established by virtue of the fact that favors given are repaid by favors received in return. In the context of the repeated prisoner's dilemma, the tit-for-tat strategy provides a simple and direct instantiation of the principle of reciprocal support \cite{axelrod1981}.

Empirical studies in humans and other primates have generally evaluated the consistency of observed patterns of interaction with the theory of reciprocity by measuring the degree to which help given is correlated with help received across pairs of individuals or families within a community (often controlling for other factors, such as kinship); this metric is typically termed ``contingency'' in the anthropological literature \cite{gurven2006} or ``reciprocity'' in the social network literature \cite{recip-note,rao1987measures,wasserman1994}. Field studies of human and non-human primates have tended to yield positive estimates of contingency in networks of directed sharing, grooming and support \cite{jaeggireview,gilby2012}; high levels of reciprocity are also a common feature of networks analyzed in the social network literature \cite{rao1987measures,jana2013reciprocity}. In either case, levels of contingency/reciprocity have shown considerable variation depending on local circumstance and the type of interaction examined \cite{gurven2006,jaeggireview,gilby2012,rao1987measures,wasserman1994,jana2013reciprocity}.

The time scales over which reciprocity may operate in natural settings remain an open area of research. On the one hand, if a simple tit-for-tat-like rule takes into account only one or a few recent interactions, feedback between an individual's actions and their partners' responses should be observable within a relatively short span of time (sufficient to capture multiple opportunities for interaction). On the other hand, decisions to lend or withhold support may take into account a pair's history of interaction over relatively longer spans of time, making their conditionality hard to detect within the space of a single observational study. Alternatively and complementarily, long-term dynamics of partner choice and selective retention may establish the conditions for reciprocity without short-term accounting of goods and services exchanged \cite{silk2003, hruschka2010, gilby2012}.

The time scales of reciprocity have proven difficult to characterize empirically. Given constraints on data collection in ethnographic settings, most tests of contingency in humans compare interaction rates summarized at a single point in time \cite{kasper} or rates that are averaged or summed over the period of sampling \cite{hames2007,gurven2004}. In one approach, Gurven's analysis of Ache food sharing showed that rates of giving to a particular family in one period correlated positively with rates of receiving from that family in the subsequent period \cite{gurven2006}. While studies of captive primates have produced some evidence for short-term contingent reciprocity \cite{hemelrijk1994,melis2008,koyama2006,jaeggi2012}, results have tended to be weak, leading researchers to suggest that the mechanisms supporting cooperative dyadic relationships may tend to operate over durations of time that exceed the span of most longitudinal studies \cite{schino2009,gilby2012,frank2009,gomes2009,Sabbatini2012,Tiddi2011,jaeggi2012}. 

Additionally, there is an open debate about the expected relationship between kinship and reciprocity in naturalistic settings. On the one hand, Axelrod and Hamilton \cite{axelrod1981} and subsequent theoretical \mbox{models \cite{mcelreath2007,mcglothlin2010interacting,akccay2012behavioral}} have predicted a positive synergy between kinship and reciprocity. Relatedness may facilitate the establishment of reciprocal exchange by increasing the effective benefits of initial altruistic acts, or increasing the probability that cooperative strategies find each other \cite{axelrod1981,mcelreath2007,allenarave2008}. Compared to non-kin, kin may often face lower transaction costs in forming reciprocal relationships, due to greater proximity, familiarity or the likelihood of repeated interaction \cite{allenarave2008}. 

If, on the other hand, kinship and reciprocity act as strategic substitutes, cooperation may be maintained by mainly altruistic motives among kin, but reciprocity among non-kin. When statistically modeling cooperative relationships in field settings, a negative interaction between kinship and contingency may also capture inequality/asymmetries within reciprocal relationships driven by the directionality of kin-based altruism (e.g., due to age or differential productivity). Empirical results on this \mbox{topic (\cite[]{kasper,allenarave2008,hooper2011,nolin2011kin}, }current study) have been mixed, which may reflect true cross-case variation in the degree to which kinship and reciprocity are complementary or substitutable means for stabilizing \mbox{social bonds.}

While generosity motivated by kinship or dyadic reciprocity implies that interactions should be patterned according to dyadic characteristics ({\em i.e.}, degree of relatedness, rates of favors returned), other evolutionary processes may support patterns of behavior that are not so clearly resolved at the dyadic level. In systems of indirect reciprocity, competitive altruism or generosity signaling \cite{nowak2005evolution,macfarlan2012competitive,gurven2000wonderful}, individuals extend help to others with the expectation of the return of benefits, not from the recipient \emph{per se}, \mbox{but from} other members of the community who become aware of the act and adjust their behavior toward the donor accordingly (see, also, Trivers' discussion of generalized altruism in \cite{trivers1971}). Accounts of ``strong reciprocity'' and cultural group selection have suggested that unconditional altruism and behaviors reinforcing prosociality (such as peer monitoring and punishment) may be favored by group or multi-level selection \cite{bowlesgintis2010,bowlesgintis2011}. 

In such situations where giving is diffuse and generalized, one would not expect strong correlations between the rates of giving and receiving within dyads, nor systematic heterogeneity in the quality of dyadic relationships (except that which is produced as a function of spatial proximity or other exogenous factors affecting rates of association). Similarly, prosocial behavior motivated by the benefits of signaling one's wealth, condition or quality \cite{smith2000turtle,gintis2001costly} generally predicts stronger patterning of behavior according to donor characteristics, rather than dyadic characteristics, such as kinship or return generosity.

\subsection{Current Aims and Approach}

The current study explores the stationary and dynamical patterns of interaction within one social network over a sixteen-week window with respect to kinship and reciprocity. Motivated by the theoretical and empirical research outlined above, we ask: to what degree is there stable heterogeneity in the intensity of dyadic social relationships (as opposed to equal mixing or generalized giving within the sample), controlling for spatial proximity? To test predictions from the theories of kin selection and dyadic reciprocity, we evaluate the degree to which this heterogeneity is associated with the degree of relatedness and patterns of reciprocation across dyads. We further ask whether there is an observable signal of direct, event-dependent reciprocation of hosting between dyads within the time scale of the study. In order to shed light on the empirical relationship between kinship and dyadic reciprocity, we ask whether and how the stationary (long-term) or dynamic (short-term) signals of reciprocity differ between related and unrelated dyads. 

For the statistical evaluation of these questions, we employ a mix of quantitative methods. Multi-level regression models are utilized to analyze the structure of stationary patterns of interaction over the sample period. We also introduce a new method for the evaluation of conditional, non-stationary or time-specific reciprocity effects, drawing on the formalism of probability calculus. This analysis builds on methods for the analysis of network dynamics developed in \cite{dedeo1,dedeo2}. 

\section{Methods}
\vspace{-12pt}
\subsection{Population and Field Setting}

The Tsimane' are an indigenous population of forager-horticulturalists living in the Beni region of northeastern Bolivia. There are currently roughly 11,000 Tsimane' residing in 90 villages in the forested region between the foothills of the Andes and the wetland-savannas of the Llanos de Moxos. The Tsimane' practice a subsistence economy of hunting, fishing, collecting and horticulture; some families in villages with access to the modern Bolivian market also engage in wage labor or cash crop rice. \linebreak It is typical for multiple, often closely-related, nuclear families to reside together in residential clusters, within which food and labor are regularly shared. Individuals and families commonly cooperate in hunting, fishing and horticultural field labor ({e.g.,} clearing and harvesting fields). Villagers also rely on each other for support during periods of illness, injury or loss \cite{hooper2011,gurven2012womb}.

Among the Tsimane', manioc beer is made by cooking and fermenting sweet manioc (sp. \emph{Manihot esculenta}) over a period of several days. While men and women both contribute to cultivating manioc, preparing the beer is almost exclusively the activity of women. After harvesting, the manioc is peeled, sliced and boiled in water for several hours in large metal pots over open fire, with intermittent stirring. The viscous mixture is then poured into wooden bins to ferment. Pieces of boiled manioc are masticated and reintroduced to the brew, which is fermented for a period of 1--6 days. Ground maize is sometimes added to increase alcohol content, which is likely to fall in the range of 0\%--4\% \cite{mowat}. The home-brew is consumed from a hollowed gourd within the household and at social gatherings and \emph{fiestas}: ``when offering \emph{chicha}, all those present are served from the same gourd in a rotating fashion, to refresh them and to satiate their stomachs, until the last drop is finished'' \cite{iamele}. Sweeter, usually less fermented \emph{chicha} is also made from maize or bananas. 

\subsection{Data Collection and Preparation}

Manioc beer network data were collected in 2007 in one Tsimane' village ({population}
 198 individuals; 35~nuclear families) located on the R{\'\i}o Maniqui, downriver from the town of San Borja, Bolivia. Over a sample period of 16 weeks, interviews were conducted in the Tsimane' language by {P.L.H.} or one of two Tsimane' research assistants with each family roughly twice per week. These interviews recorded the family's reports of preparing, drinking and serving manioc beer during the preceding two days. 

For analysis of the manioc beer dataset as a network, nuclear families constitute nodes, while the daily frequencies of hosting constitute the weight of directed edges that connect them. The sample of \mbox{54,218 possible }directed dyadic interactions includes one observation for each day that a given family was at risk of being observed hosting another given family (because either the potentially hosting or hosted family completed an interview covering that day). Given an observed tendency for under-reporting, observation days for which two families' reports disagree on attendance were reconciled in favor of the positive report. Eleven instances in which hosts and guests reported divergent dates for what was likely to be the same event were randomly reconciled in favor of one of the two reports. 

Kinship was estimated on the basis of genealogical data (census and demographic interviews) collected by the Tsimane' Health and Life History Project throughout 2006--2007 \cite{gurven2007mortality}. The genetic relatedness between a given pair of families, $i$ and $j$, was calculated as the mean relatedness of each member of family $i$ with each member of family $j$ on the basis of $\ge$3 generations of genealogical data using the kinship2 package in R \cite{kinship2,rlang}. Distances between families were calculated according to the spherical law of cosines using household locations recorded with a handheld GPS unit. The hosting and kinship networks were visualized using the igraph package in R, with node layouts optimized by the Fruchterman-Reingold algorithm \cite{igraph}.

\subsection{Multi-Level Regression}

For the stationary analysis of hosting patterns between families, multi-level logistic regression models predicting the probability that one family hosted another on a given risk day, were estimated using the glmer function in the lme4 package in R \cite{lme4,model-note}. These models take the form: 
\begin{equation}
\textrm{ln} \left( \frac{ P(\textrm{``$i$ hosts $j$'' at time $t$})}{1- P(\textrm{``$i$ hosts $j$'' at time $t$}) } \right) = \alpha _{i} + \alpha _{j} + \alpha _{ij} +\mathbf{ X \beta }
\label{logistic}
\end{equation}


\noindent
Stationary heterogeneity in the patterning of interactions across families and dyads was modeled with the inclusion of random effects, $\alpha _{i}$, $\alpha _{j}$ and $\alpha _{ij}$, for the identities of hosts, guests and host-guest dyads, respectively. The use of random effects allows for the partitioning of variance between multiple levels of aggregation within the data. The variance of the random effects in these models indicates the share of variance unexplained by the model's predictors ($ \mathbf{ X \beta }$), which is attributable to higher-order groupings, in this case, host, guest and host-guest dyad identity \cite{gelman2007data,hox2010multilevel}. 

\begin{figure}
\centering
\includegraphics[width=4in]{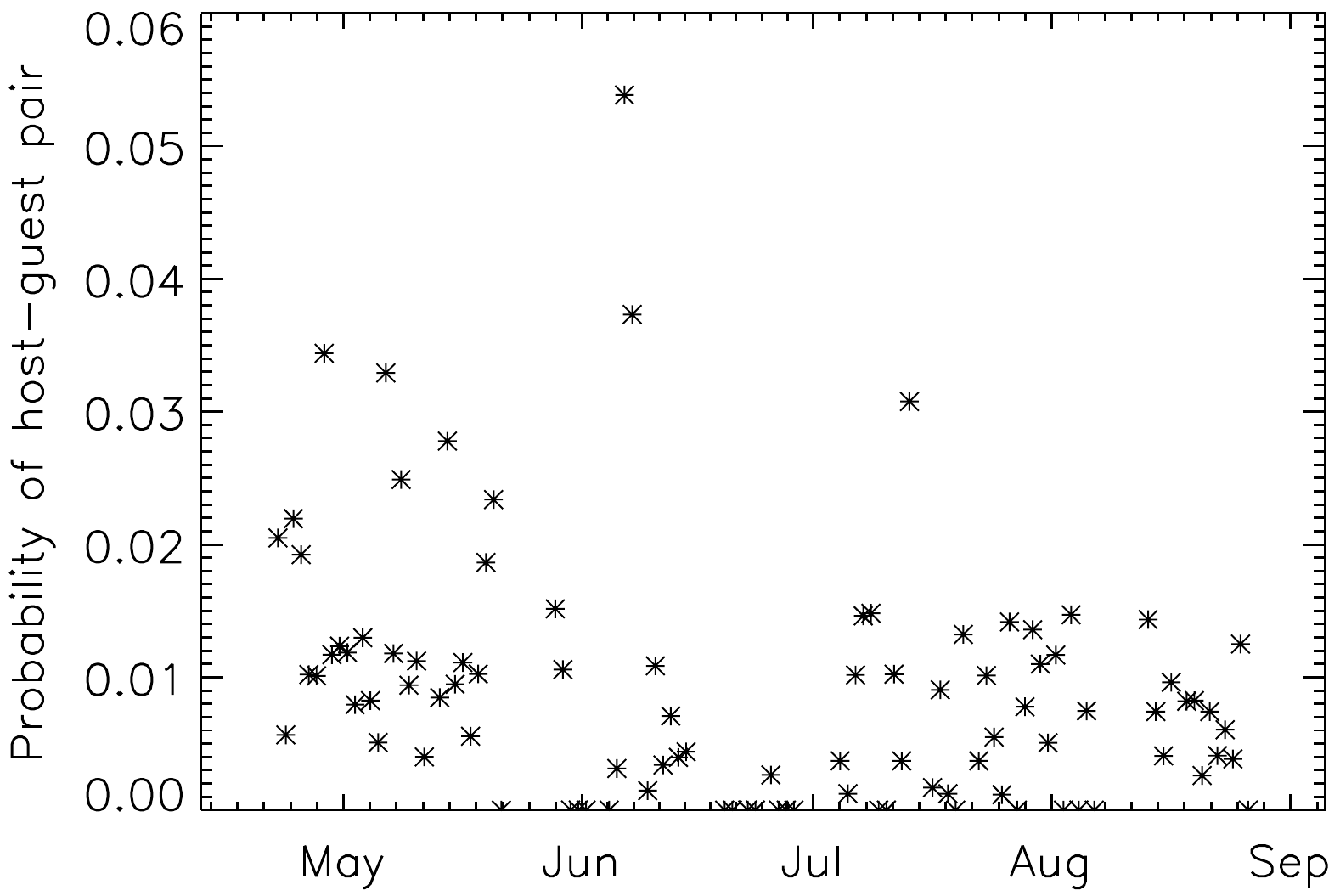}
\caption{Seasonal effects in hosting. Shown here, for the discrete sampling points, is the probability that an observed pair is a hosting pair. A lull in late June separates two more active periods; the May period is overall the most active, followed by the July and \mbox{August season}.}
\label{season-data}
\end{figure}

To account for differential under-reporting by guests relative to hosts, a fixed effect was included that flagged potential dyadic events that were covered in the interviews of potential guests, but not hosts. Changes in the frequency of hosting over time (see Figure~\ref{season-data}) were modeled with the inclusion of fixed effects for each week of observation. The ability of family-level characteristics of hosts and guests---the mean age of household heads and number of family members---to explain variation in rates of hosting and attendance was evaluated on the basis of minimization of Akaike's Information Criterion (AIC). 

The ability of spatial proximity, kinship and stationary reciprocity ({\em i.e.}, contingency) to explain variation in the probability of family $i$ hosting family $j$ was evaluated by inclusion of predictor variables representing: (1) the natural logarithm of distance between households (km); (2) the mean genetic relatedness of each dyad; and (3) the mean rate that family $j$ hosted $i$ over the sample period. The potential for different patterns of reciprocation depending on kinship status was evaluated with inclusion of an interaction between relatedness and reciprocal hosting.

We recognize that correlations between the rates that $i$ hosts $j$ and $j$ hosts $i$ may arise for a number of reasons. For example, $i$ may visit $j$ and \textit{vice versa} simply because $i$ and $j$ live in close proximity or because $i$ and $j$ are closely related. Correlations due entirely to such factors would not provide strong evidence for the kind of contingent reciprocity described in the evolutionary literature; the presence, however, of a significant relationship controlling for ``confounding'' third factors, such as distance or kinship, provides stronger support for the hypothesis that relationships are maintained in part by the logic of reciprocity ({\em i.e.}, that the generosity of $i$ to $j$ is encouraged by the generosity of $j$ to $i$, and \textit{vice versa}). It is also worth recognizing that the use of spatial proximity as an independent control in tests of the effects of kinship and reciprocity is conservative, in the sense that household locations are endogenous choices, which are made, in part, with respect to the location of one's kin and close social partners; this is especially true among the Tsimane', who exhibit high residential mobility and are unconstrained by private ownership of land. Twenty-two percent of the variance in distance between families is explained by genetic relatedness in the current dataset.

While random effects account for some aspects of the non-independence of outcomes across observations, conventional significance tests for models of network data may be biased, due to other aspects of \mbox{non-independence} in the data structure. Significance values for estimates of fixed and random effects were therefore bootstrapped by comparing observed values against the distribution of values produced from resampled data in which hosting outcomes (0 or 1) were randomly shuffled across~observations.

\subsection{Time-Series Analysis}

In addition to our analysis of stationary patterns, we also look for \emph{dynamical} signatures of reciprocity. We look for evidence, at all timescales---days to months---available in our data, for causal and \mbox{non-stationary} relationships between families mediated by hosting. To do this, we construct a general model for pairwise interactions, which we refer to as ``conditional-action'' reciprocity.

Informally, conditional-action reciprocity is when family $j$ ``returns the favor'' of being hosted by family $i$, by themselves hosting a member of family $i$ at a later date. Formally, we consider the ratio of the conditional probability to the base rate:
\begin{equation}
R_{ij}(\Delta t) = \frac{P(\textrm{``$j$ hosts $i$'' at time $t+\Delta t$}~|~\textrm{``$i$ hosts $j$'' at time $t$})}{P(\textrm{``$j$ hosts $i$''})}
\label{recip}
\end{equation}
or, in other words, the fraction by which the probability of $j$ hosting $i$ is increased given a prior hosting of $j$ by $i$.

The definition of Equation~\eqref{recip} has intuitive appeal as a way to track the underlying mechanisms by which reciprocity might develop and be maintained over time. Care, however, must be taken in systems where patterns of pro-social behavior are connected to long-standing social expectations. Imagine, for example, that families $i$ and $j$ have established a practice where, on average, each hosts the other five times a month. Each family has a constant probability of hosting ($\approx$1/6 per day), which is unchanged by the occurrence of a hosting event by the other. Such a pattern would thus be invisible to the measure above, despite the fact that it might well be part of a social practice maintained by the norms of~reciprocity.

Interruptions of the pattern established by $i$ and $j$ may lead to a detectable signal, as might the (reciprocated) decision by one family to increase the frequency with which they host the other. Such essentially non-stationary effects are necessary to lead to a detectable change in the conditional probabilities underlying the definition in Equation~\eqref{recip}. Conditional-action definitions of reciprocation have the paradoxical effect that they lead to zero signal, not only in non-reciprocal populations, but also in highly pro-social systems, where patterns of reciprocation are \emph{stable} and \emph{unchanging}.

In analogous cases in the non-human world, interventions are possible, in which variables are altered by external manipulation (the ``do'' operation of~\cite{pearl}). This allows an experimenter to interrupt stable patterns of reciprocity---by, say, preventing the commission, or undoing the omission, of a favor---to detect an underlying conditional law. For reasons of both ethics and practicality, however, data in human systems are often observational, as opposed to interventional, in nature. In our case, we only observe what people have done, rather than asking them to behave differently. This makes dynamical and causal relationships more difficult, though not in all cases impossible, to detect. In particular, it requires more sophisticated attention to both estimation methods and to null model design, which we address further in Section~\ref{hier}

\subsubsection{Estimating Conditional-Action Reciprocity}

The data consist of samples taken on discrete days, concerning the events of the two previous days. Sampling was incomplete; we account for all sampling effects by means of an observation window, $o_{ij}(t)$. Here, $o_{ij}(t)$ is unity if the hosting event ``$i$ hosts $j$ at time $t$'' was \emph{potentially} observable (if either the host or the guest were interviewed). The actual hosting events we represent by $h_{ij}(t)$, where $h_{ij}(t)$ is unity if $i$ was indeed observed to host $j$ at time $t$, and zero otherwise. By definition, if $h_{ij}(t)$ is non-zero for a particular pair and time, so is $o_{ij}(t)$.

Given these definitions, any particular $R_{ij}$ pair can be estimated, correcting for observational windowing effects:
\begin{equation}
\hat{R}_{ij}(\Delta t) = \left(\frac{\sum_{t,t^\prime; t^\prime-t=\Delta t}h_{ij}(t)h_{ji}(t^\prime)}{\sum_{t,t^\prime; t^\prime-t=\Delta t}o_{ij}(t)o_{ji}(t^\prime)}\right) \left(\frac{\sum_{t,t^\prime; t^\prime-t=\Delta t}o_{ij}(t) o_{ji}(t^\prime)}{\sum_{t,t^\prime; t^\prime-t=\Delta t}h_{ij}(t)o_{ji}(t^\prime)}\right) \left(\frac{\sum_{t,t^\prime; t^\prime-t=\Delta t}o_{ij}(t)o_{ji}(t^\prime)}{\sum_{t,t^\prime; t^\prime-t=\Delta t}o_{ij}(t)h_{ji}(t^\prime)}\right)
\label{rest}
\end{equation}
where we use the rule:
\begin{equation}
P(A|B) = \frac{P(A,B)}{P(B)}
\label{prob1}
\end{equation}
and the maximum-likelihood estimator:
\begin{equation}
P(A) = \frac{N(A)}{N}
\label{prob2}
\end{equation}
where $N$ is the number of observations and $N(A)$ the number of observations where event type A was observed to obtain. As the amount of data becomes large, $\hat{R}$ becomes an increasingly good estimator of~$R$; in particular, as:
\begin{equation}
\left(\sum_{t,t^\prime; t^\prime-t=\Delta t} h_{ij}h_{ji}\right) \rightarrow \infty
\end{equation}
then:
\begin{equation}
\hat{R}_{ij} \rightarrow R_{ij}
\end{equation}

In the limit of infinite data, and arbitrarily long time-streams, a system without dynamical, conditional-action reciprocity will measure $R_{ij}$ for each $i$-$j$ pair at its expectation value of one. Because the measurement of reciprocity in any particular $i$-$j$ pair is expected to be extremely noisy, we consider $R_{ij}$ estimates averaged over pairs drawn from sets of interest. We focus on three groups: all pairs with observed reciprocity, all close kin pairs with observed reciprocity and all distant and non-kin pairs with observed reciprocity. We renormalize the observed values of $R_{ij}$ by reference to the expectation for that particular pair under the null model of interest, so that we report:
\begin{equation}
\hat{R}(\Delta t) = \left\langle \frac{\hat{R}_{ij}(\Delta t)}{\langle\hat{R}_{ij}(\Delta t)\rangle_\textrm{null}} \right\rangle_{ij\in C}
\label{max-l-average}
\end{equation}
where averaging over realizations of a particular null model is written $\langle\ldots\rangle_\textrm{null}$ and averaging over sets of $i$-$j$ pairs is written $\langle\ldots\rangle_{ij\in C}$; the class, $C$, might be ``all pairs'', ``close kin pairs'' or ``not close kin pairs''.

\subsubsection{A Three-Level Hierarchy of Null Models for Conditional-Action Reciprocity}
\label{hier}

Null models are a standard tool for determining to what information an underlying mechanism needs to have access. Use of a \emph{hierarchy} of null models (with increasingly strict nulls forcing the allowed behavior closer and closer to the observed data) allows us to gather information for or against different kinds of underlying effective mechanisms. Previous studies of behavioral systems in non-human primates have used null hierarchies to determine the role of pairwise \textit{vs}. higher-order relationships in conflict~\cite{dedeo1} and to determine the relative role of exogenously \textit{vs}. endogenously-driven decision-making in conflict management~\cite{dedeo2}.

Here, we use nulls to compare measurements of $\hat{R}_{ij}$ to those measured on simulated datasets that match (on average) various stationary properties of the data while selectively destroying correlations. This allows us to determine both effect \emph{size} (how much larger $\hat{R}_{ij}$ is than unity, for example---a direct measure of the level of conditional-action reciprocity in the system) and effect \emph{significance} (whether or not an effect of a particular size could have occurred by chance in a model without an underlying condition-action mechanism). These nulls amount to large collections of simulated datasets. We measure the same quantities (such as average $\hat{R}_{ij}$) on this simulated data and determine whether or not we can establish evidence for mechanisms ``over and above'' those allowed for by the nulls.

We consider three null models of increasing strictness. The mechanisms underlying all three nulls are time-independent: {\em i.e.}, they contain no conditional-action mechanisms. Each observation period is an {independent and identically distributed (i.i.d.)} 
~sampling from mechanisms defined by what features of the original data they hold constant.

The first null we call the \emph{homogeneous null}. In this case, we preserve, on average, each family's probability of hosting a party and the distribution of the sizes of the parties they do host. In particular, we preserve:
\begin{equation}
P(\textrm{``$i$ hosts a party''})
\end{equation}
and (by means of a sufficient statistic for the binomial distribution):
\begin{equation}
P(\textrm{``$i$ hosts a party of size $N$''})
\end{equation}
This null preserves many features of the data. Families that host many parties do so (on average) in this null and with the same size distribution. However, the preferences that a host might have to host one family over another are erased. In this model, parties are ``pure sociability''---they reflect only an individual family's desire to share resources with the larger community.

The second, stricter, null is the \emph{kin-heterogeneous null}. In this case, we allow hosting preferences to vary, but require them, for each host, to be identical for all families with the same mean $r$. In this model, parties are reflections of long-standing, stationary preference that particular hosts have for kin groups. These preferences are allowed to be asymmetric, meaning that $P(\textrm{``$i$ hosts $j$''})$ and $P(\textrm{``$j$ hosts $i$''})$ can~differ.

The third, and most strict, null is the \emph{full-heterogeneous null}. In this case, we preserve, on average, \linebreak all host-guest pairwise probabilities; in particular, the probability of observing a particular host-guest pair, $P(\textrm{``$i$ hosts $j$''})$ is preserved on average. In this model, parties are reflections of long-standing, stationary preferences between particular host-guest pairs. As in the kin-heterogeneous null, these preferences are allowed to be asymmetric.

These three nulls specify how to construct a set of probabilities for hosting, $P(\textrm{``$i$ hosts $j$''})$, which can then be sampled, in an i.i.d. fashion, to produce simulated data, which is then windowed by the actual observation structure and from which the conditional-action reciprocity term, $\hat{R}_{ij}(\Delta t)$, can be estimated. We make, for the sake of accuracy, two minor corrections to this account---a non-stationary correction for seasonal effects and a stationary correction for differences in how guests {\em vs.}~hosts report information to the observers; these corrections are described in detail in the Appendix, Section~\ref{app}.

\section{Results}

We now summarize a number of characteristics of the network and report our results for kinship and stationary and dynamical
reciprocity.

\begin{figure}[H]
\centering
\begin{tabular}{cc}
\includegraphics[width=3.0in]{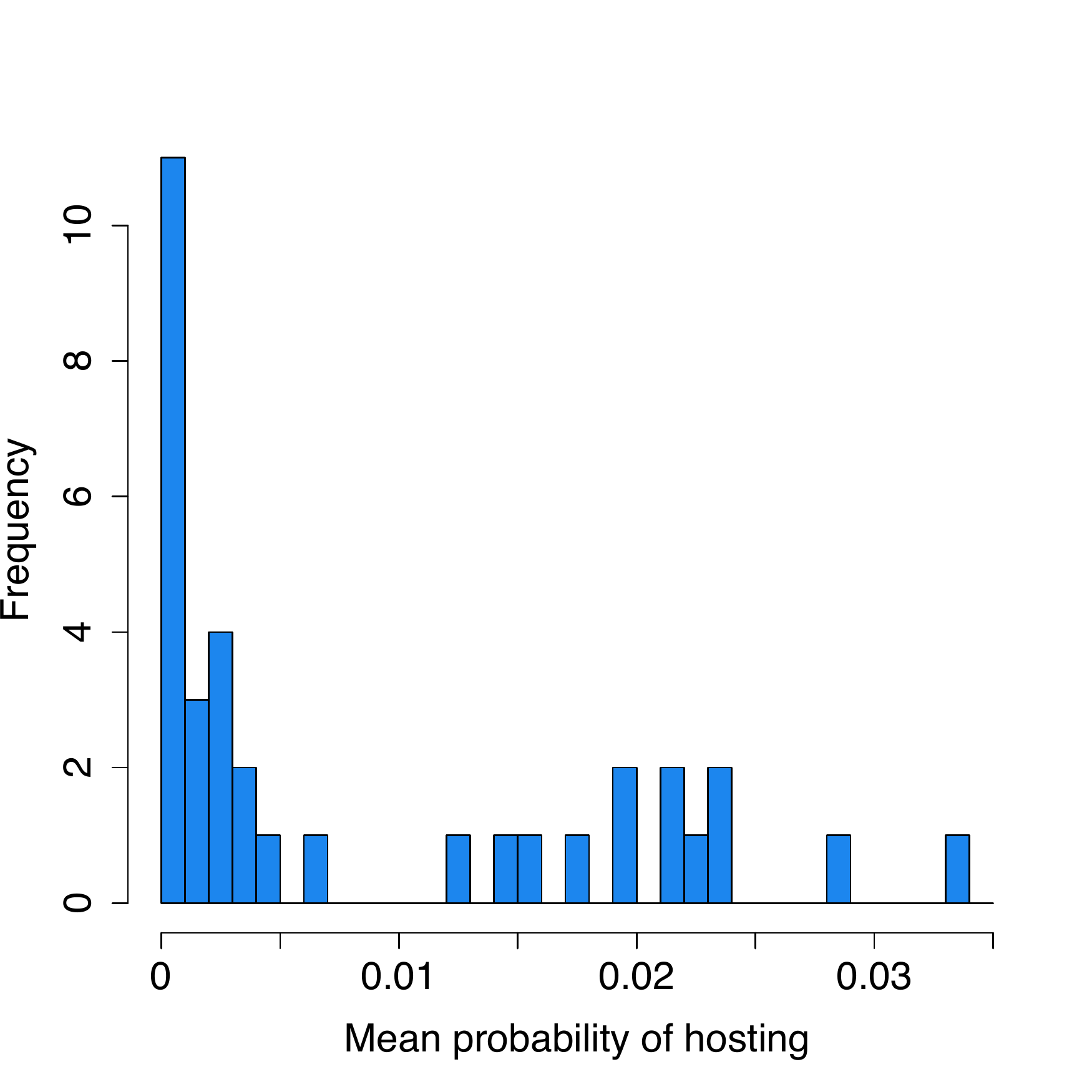} &\includegraphics[width=3.0in]{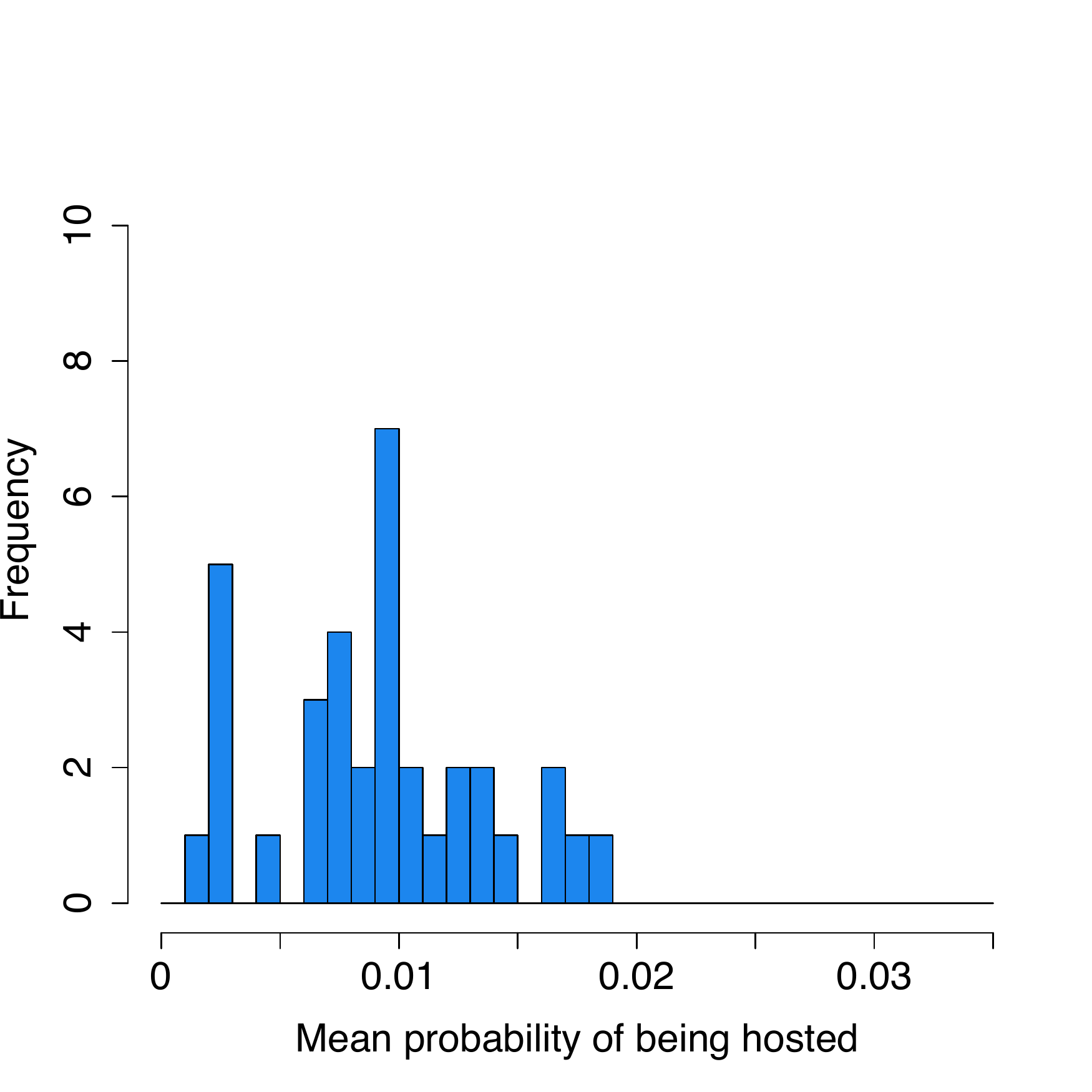} \\
(\textbf{a}) & (\textbf{b}) \\
\end{tabular}
\caption{Frequency distributions of mean daily probability of (\textbf{a}) hosting and (\textbf{b}) being hosted for any given family each sample day for 35 families.}
\label{host-dist}
\end{figure}

\begin{figure}[H]
\centering
\begin{tabular}{cc}
\includegraphics[width=0.45\textwidth]{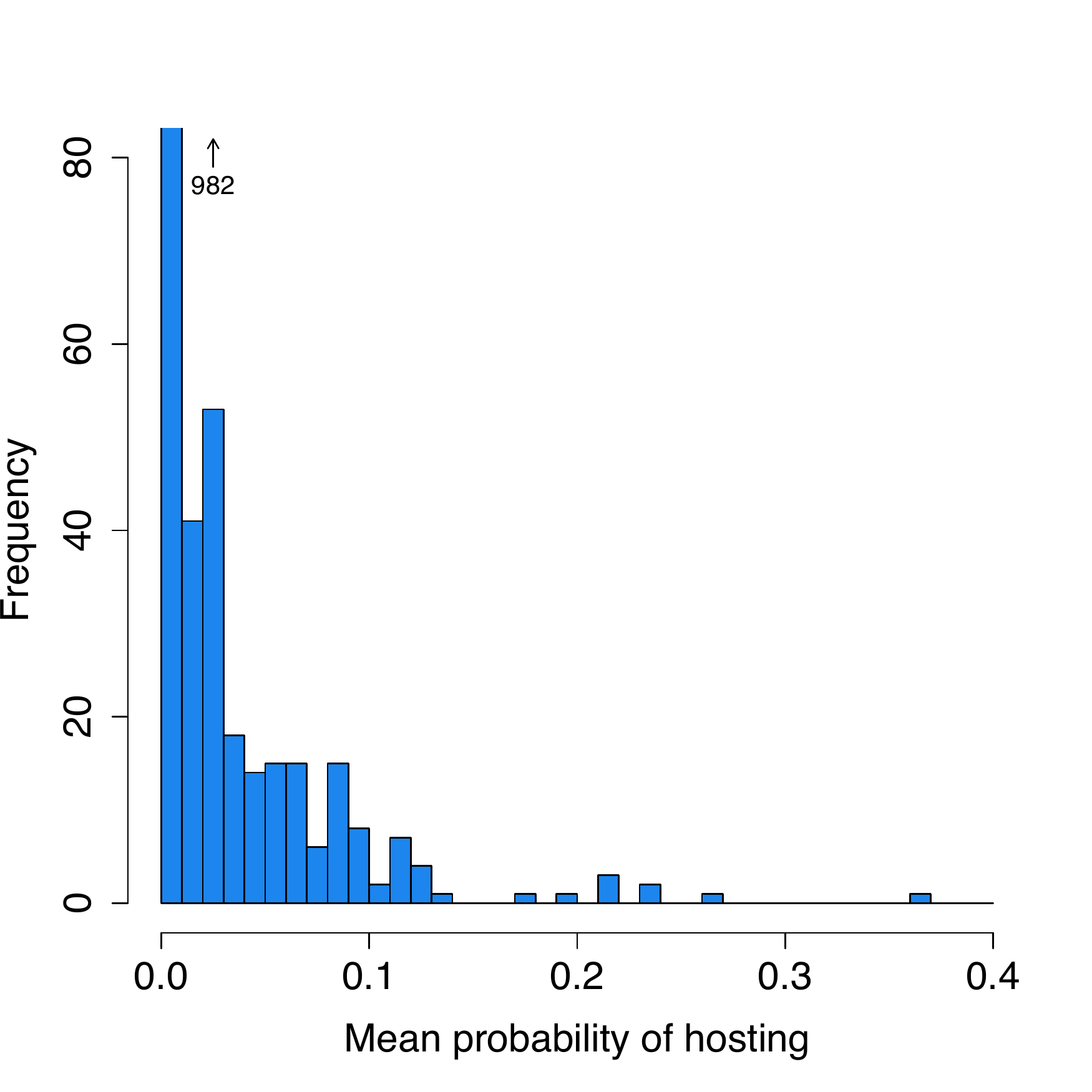} & \includegraphics[width=0.45\textwidth]{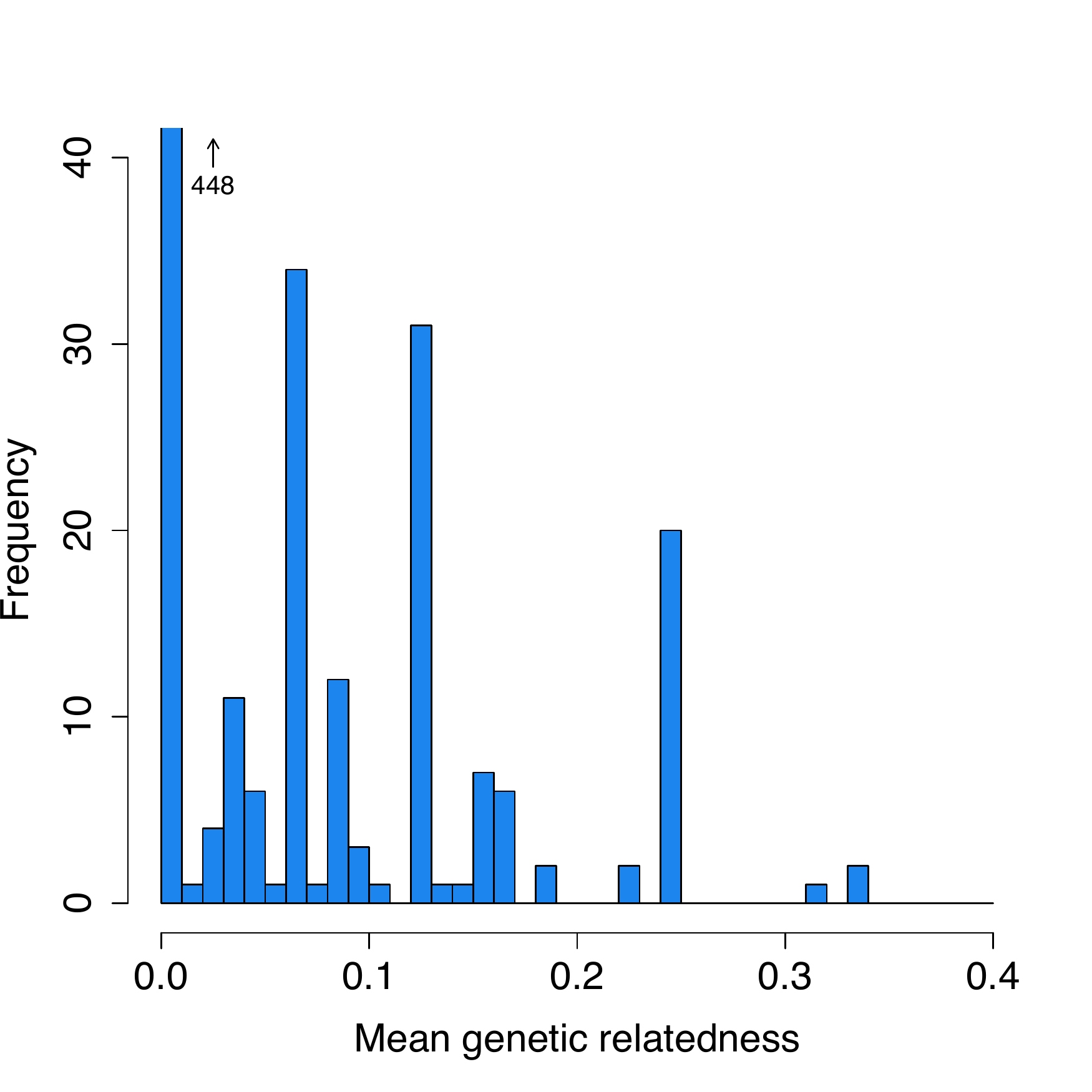} \\
(\textbf{a}) & (\textbf{b})\\
\end{tabular}
\caption{Frequency distributions of the (\textbf{a}) mean daily probability of one given family hosting another and (\textbf{b}) mean genetic relatedness between families for 1190 \mbox{host-guest dyads.}}
\label{r-dist}
\end{figure}

\subsection{Summary Statistics}

The mean daily probability of one family hosting another given family measured over the sample period is 0.926\% (502$/$54,218 host-guest risk days). The distributions of the frequency of hosting and being hosted across families are given in Figure~\ref{host-dist}a,b, respectively. It is clear that while most families attend \emph{chicha} events, only some families host them frequently; as a result, the variance in the rate of hosting is roughly five times that of the rate of being hosted. 

The distribution of the weights of host-guest relationships over the sample period is given in Figure~\ref{r-dist}a. At least one hosting event was observed for 17\% (208$/$1190) of possible host-guest dyads. The distribution of mean genetic relatedness between families is given in Figure~\ref{r-dist}b. For the genealogical data available (which may omit more distant ancestry), the mean relatedness of families in this community is 0.0293. Roughly 4\% (23$/$595) of family dyads are related with mean $r\geq$ 0.25 (principally parent-offspring and sibling dyads); 21\% (124$/$595) are related with 0 $<r<$ 0.25 (secondary and more distant kin); while 75\% (448$/$595) are unrelated. 

\begin{figure}
\centering
\includegraphics[width=3.4in]{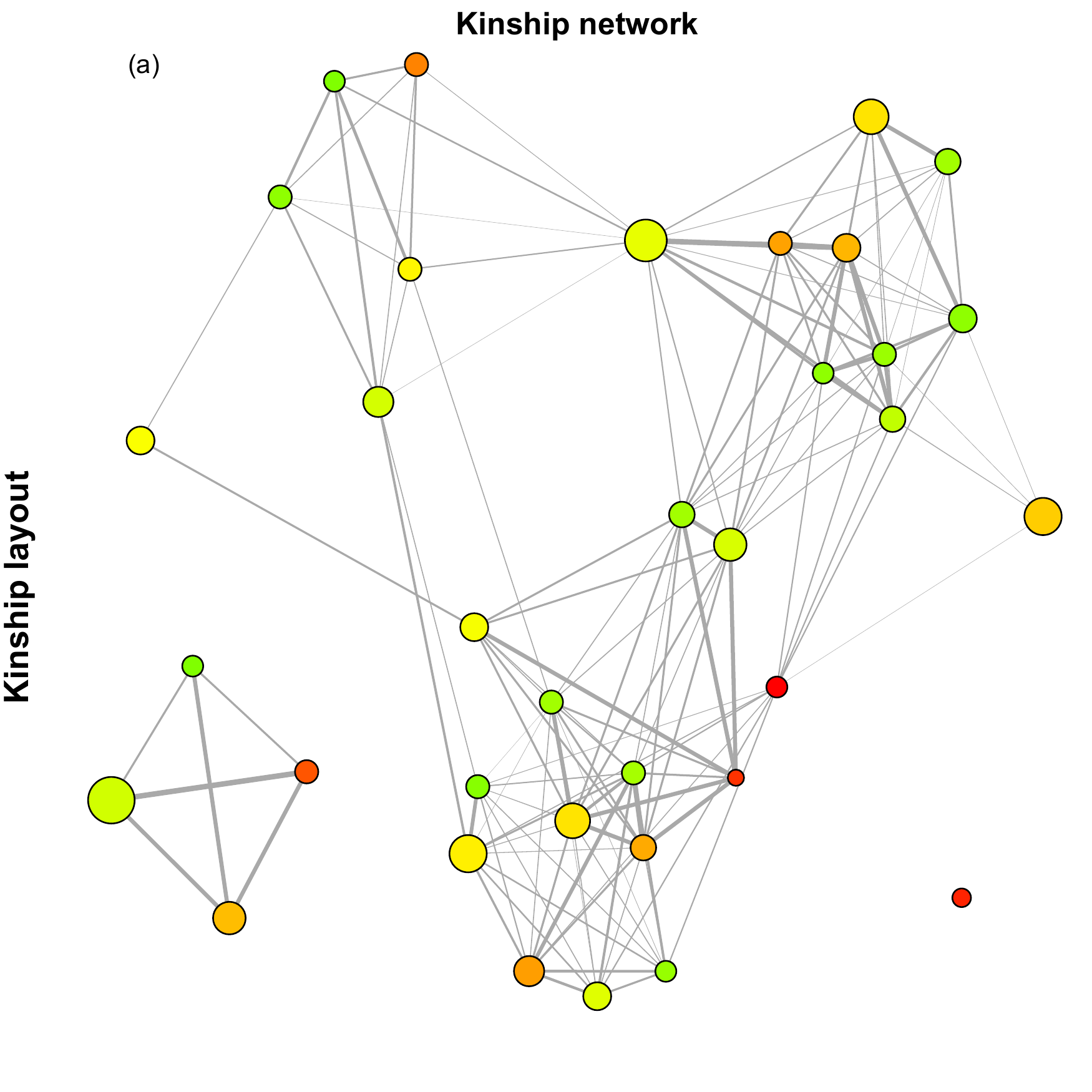}
\includegraphics[width=3.4in]{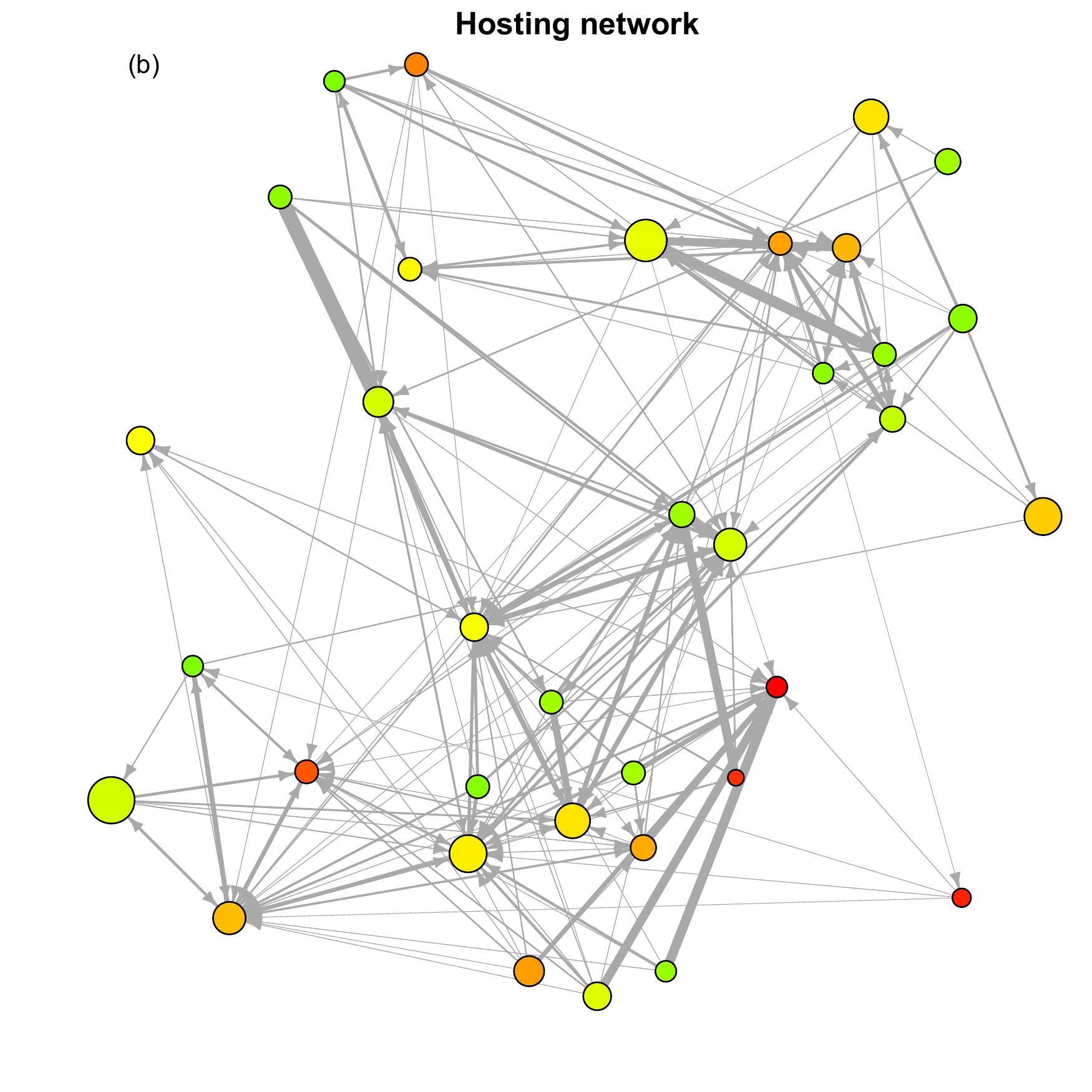}
\includegraphics[width=3.4in]{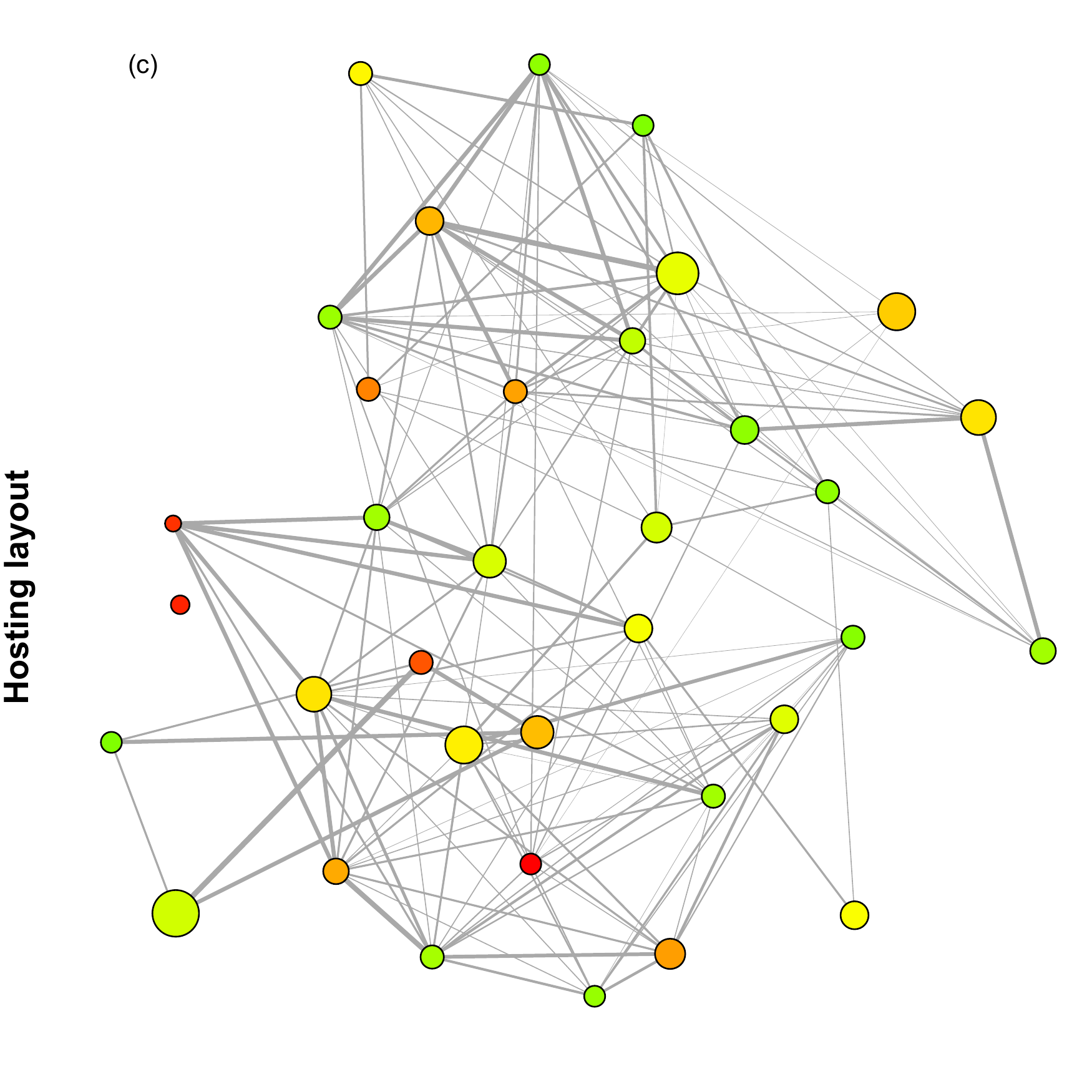}
\includegraphics[width=3.4in]{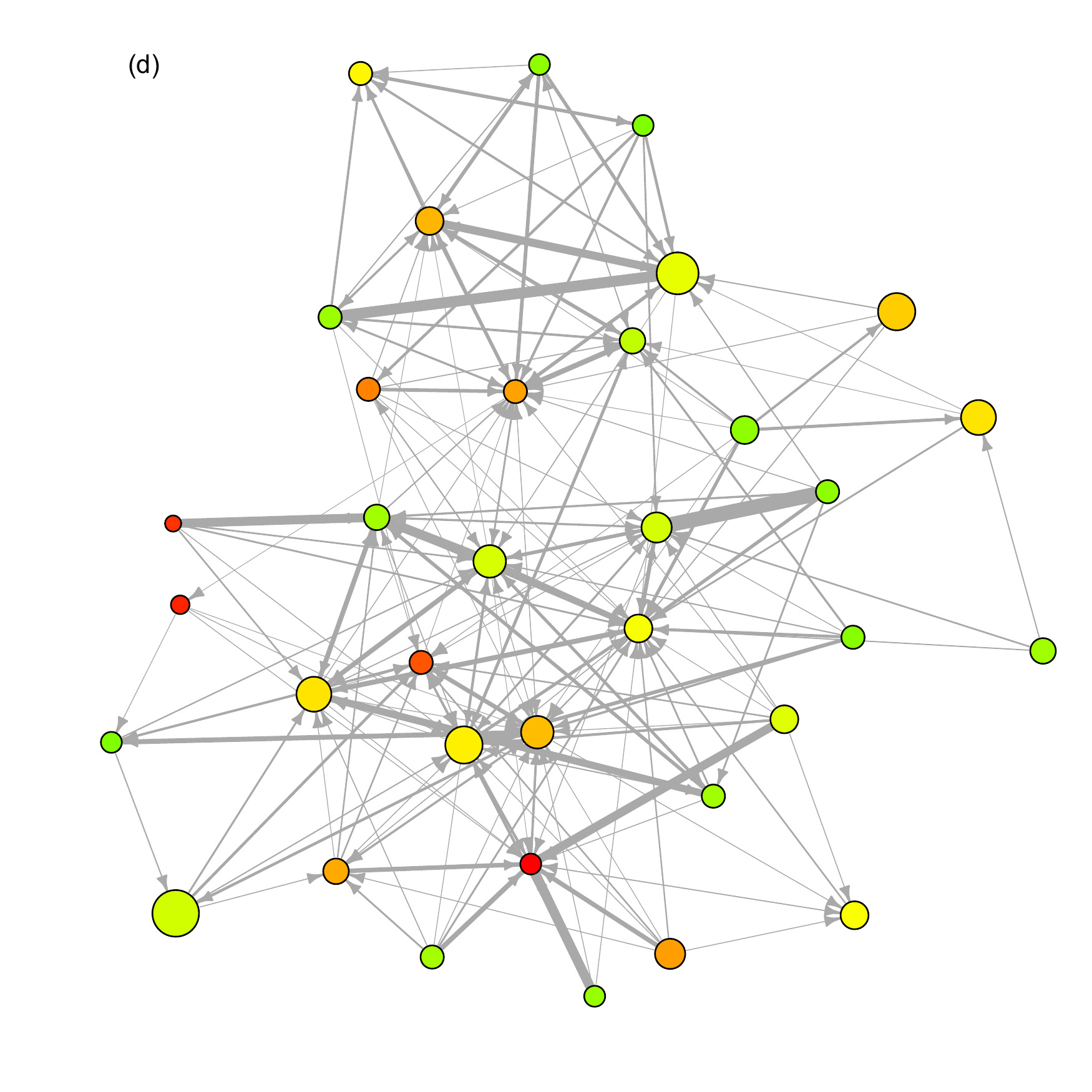}
\caption{Networks of kinship and manioc beer hosting. Panels (\textbf{a}) and (\textbf{c}) represent the kinship network, with the edge width indicating mean genetic relatedness. Panels (\textbf{b}) and (\textbf{d}) represent the hosting network; edge widths indicate the mean hosting frequency between the two families, while arrows indicate the direction of attendance. The layouts of panels (\textbf{a}) and (\textbf{b}) are optimized based on the kinship network, while those of (\textbf{c}) and (\textbf{d}) are optimized based on the hosting network. Node color indicates mean age of household heads (ascending in age from green to red), while node size indicates family size.}
\label{four-networks}
\end{figure}

Figure~\ref{four-networks} plots the hosting and kinship networks with two different layouts: the layout of the first row of networks is set according to the kinship network, while that of the second row is set according to the hosting network. For either layout, comparison of the hosting network with the kinship network shows that while hosting tends to occur most frequently between close kin, many hosting relationships also occur between non-kin; there is also clear variation in the relative magnitude of relationships within both related and unrelated interacting dyads.

\subsection{Multi-Level Regression}
\label{MLR}

Table~\ref{mlr-table} gives the results of the multi-level regression predicting the likelihood of host-guest interactions across the sample period. Baseline parameter estimates with the inclusion of controls for the week of observation and reporting bias are given in Model A. The variance of the random effects in this baseline model quantifies the degree to which hosting is idiosyncratically distributed across hosts, guests and host-guest relationships. Confirming the impression given in Figure~\ref{host-dist}, there is substantially greater systematic variation in the propensity to host than in the propensity to attend (2.581 {\em vs.} 0.046). The variance in hosting that is attributable to the identity of specific host-guest dyads (2.733) is also significant, indicating substantial heterogeneity in the intensity of relationships between families within the community. The reporting control confirms that events covered only by guests' interviews were reported at significantly lower rates than those covered by hosts'.

Model B shows the best-fit model for hosting frequency with the inclusion of family-level predictors. The rate of hosting and attendance varies significantly with the mean age of household heads, with AIC minimization favoring quadratic and linear fits for host and guest ages, respectively. The frequency of hosting first increases with age, peaks around age 50 and declines thereafter; the frequency of attendance, on the other hand, decreases weakly with age. There is also a positive association between the frequency of hosting and family size ($\beta=$ 0.224, \textit{p} = 0.024), but the effect is not significant and does not yield improvements in AIC when age terms are also included in the model. Rates of attendance are not significantly predicted by the family size of guests ($\beta=$ 0.028, \textit{p} = 0.416)~\cite{data-note}.

Models C, D, E and F assess the extent to which proximity, kinship and the stationary rate of reciprocal hosting can account for idiosyncratic variation in the quality of host-guest relationships. Models C and D show that on their own, household distance and kinship are each able to explain 59\% (1$-$1.166$/$2.817) and 60\% (1$-$1.120$/$2.817) of the variance attributed to dyad identity in \mbox{Model B,} respectively. Included together in model E, each of these variables independently predicts hosting frequency and together explain a total of 71\% (1$-$0.824$/$2.817) of the variance attributed to dyad identity in model B.

\begin{table}[H]
\vspace{-6pt}
\footnotesize
\begin{center}
\begin{tabular}{@{}l|ccc|ccc|ccc@{}}
\hline
& & \textbf{A} & & & \textbf{B} & & & \textbf{C} & \\ \hline
\emph{Random effects} & Var($\alpha$) & & \textit{p} & Var($\alpha$) & & $p$ & Var($\alpha$) & & $p$ \\ \hline
Host identity ($\alpha _{i}$) & 2.581 & & $<$0.001 & 1.875 & & $<$0.001 & 3.262 & & $<$0.001 \\
Guest identity ($\alpha _{j}$) & 0.046 & & 0.008 & 0.018 & & 0.082 & 0.135 & & $<$0.001 \\
Host-guest dyad ($\alpha _{ij}$) & 2.733 & & $<$0.001 & 2.817 & & $<$0.001 & 1.166 & & $<$0.001 \\ \hline
\emph{Predictors} & $\beta$ & $e^ \beta $ & $p$ & $\beta$ & $e^ \beta $ & $p$ & $\beta$ & $e^ \beta $ & $p$ \\ \hline
Guest report only & $-$0.972 & 0.378 & $<$0.001 & $-$0.964 & 0.381 & $<$0.001 & $-$0.956 & 0.384 & $<$0.001 \\
Host age & & & & 0.223 & 1.250 & $<$0.001 & 0.293 & 1.340 & $<$0.001 \\
Host age$^2$ & & & & $-$0.002 & 0.998 & $<$0.001 & $-$0.003 & 0.997 & $<$0.001 \\
Guest age & & & & $-$0.009 & 0.991 & $<$0.001 & $-$0.007 & 0.993 & 0.006\\
Log distance & & & & & & & $-$1.050 & 0.350 & $<$0.001 \\ \hline
\emph{AIC
} & & 4,565 & & & 4,560 & & & 4,365 & \\ \hline \hline
& & \textbf{D} & & & \textbf{E} & & & \textbf{F} & \\ \hline
\emph{Random effects} & Var($\alpha$) & & $p$ & Var($\alpha$) & & $p$ & Var($\alpha$) & & $p$ \\ \hline
Host identity ($\alpha _{i}$) &2.778& & $<$0.001 &3.164& & $<$0.001 &3.276& & $<$0.001 \\
Guest identity ($\alpha _{j}$) &0.129& & $<$0.001 &0.111& & $<$0.001 &0.032& & 0.030 \\
Host-guest dyad ($\alpha _{ij}$) &1.120& & $<$0.001 &0.824& & $<$0.001 &0.619& & $<$0.001 \\ \hline
\emph{Predictors} & $\beta$ & $e^ \beta $ & $p$ & $\beta$ & $e^ \beta $ & $p$ & $\beta$ & $e^ \beta $ & $p$ \\ \hline
Guest report only &$-$0.965& 0.381& $<$0.001 &$-$0.954&0.385& $<$0.001 &$-$0.959&0.383& $<$0.001 \\
Host age &0.229& 1.257& $<$0.001 &0.268&1.307& $<$0.001 &0.276&1.318& $<$0.001 \\
Host age$^2$ &$-$0.002&0.998& $<$0.001 &$-$0.003&0.997& $<$0.001 &$-$0.003&0.997& $<$0.001 \\
Guest age &$-$0.007&0.993&0.008&$-$0.007&0.993&0.008&$-$0.009&0.991& $<$0.001 \\
Log distance &&&&$-$0.618&0.539& $<$0.001 &$-$0.520&0.595& $<$0.001 \\
Kinship &14.02& 1.2E6& $<$0.001&8.860&7045& $<$0.001 &8.435&4606& $<$0.001\\
Rate $j$ hosts $i$ &&&& & & &19.99&4.8E8& $<$0.001\\
Kinship $\times$ Rate $j$ hosts $i$ & & & & & & &$-$64.86&7E-29& 0.234\\ \hline
\emph{AIC} & &4,354&& &4,294& & &4,255& \\ \hline
\end{tabular}
\end{center}
\caption{Multi-level logistic regression models predicting the likelihood that family $i$ hosts family $j$ on a given sample day (the models include intercept and time controls, not shown).}
\label{mlr-table}
\end{table}

\vspace{-12pt}
The full model including terms for distance, stationary reciprocity, kinship and their interaction is given in Model F. This model shows that the mean rate at which family $j$ hosts family $i$ is a substantially positive predictor of the likelihood of family $i$ hosting $j$ on any given day, controlling for both spatial proximity and kinship. The coefficient on the reciprocity term indicates that a 0.01 increase in the daily rate that $j$ hosts $i$ is associated with a 22\% increase in the likelihood that $i$ hosts $j$ on a given sample day. The combination of the three dyadic predictors together account for roughly 78\% (1$-$0.619$/$2.817) of the total variance attributable to dyad identity in Model B. Model F also estimates a negative interaction between kinship and reciprocal hosting, which may suggest the following interpretation: while the baseline level of interaction is higher among kin than non-kin, the estimated marginal effect of $j$ hosting $i$ on the likelihood of $i$ hosting $j$ is greater for non-kin than kin. The interaction, however, is not statistically significant when compared against resampled null models. 

In sum, the regression results confirm the existence of stationary heterogeneity in the quality of relationships across the sample period, which is patterned according to and well explained on the basis of both kinship and stationary reciprocity, above and beyond the effects of spatial proximity.

\subsection{Time-Series Analysis}
\label{TSA}

\begin{figure}[H]
\centering
\begin{tabular}{cc}
\includegraphics[width=3in]{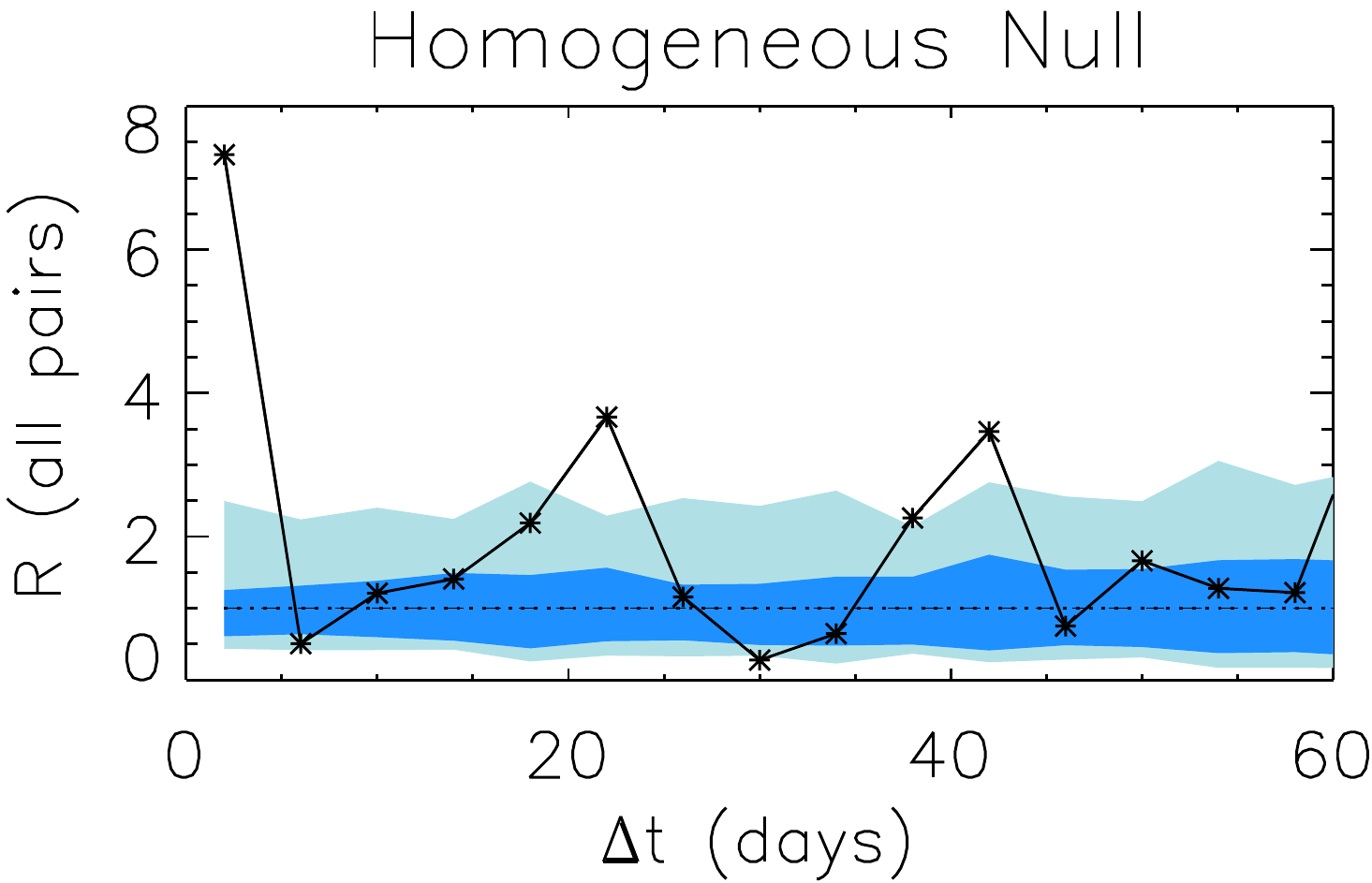} & 
\includegraphics[width=3in]{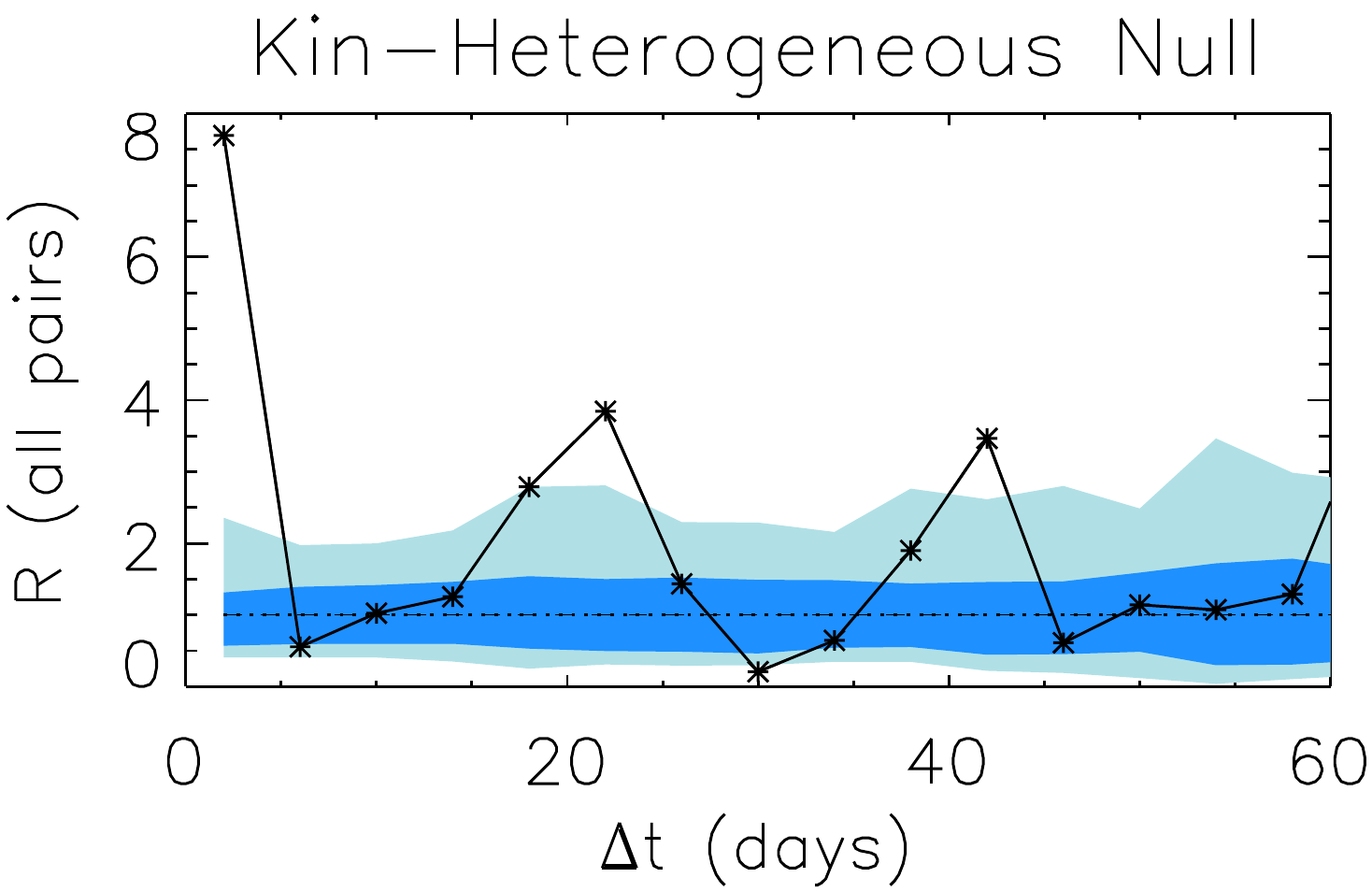} \\
({\bf a}) & ({\bf b}) \\
\multicolumn{2}{c}{
\includegraphics[width=3in]{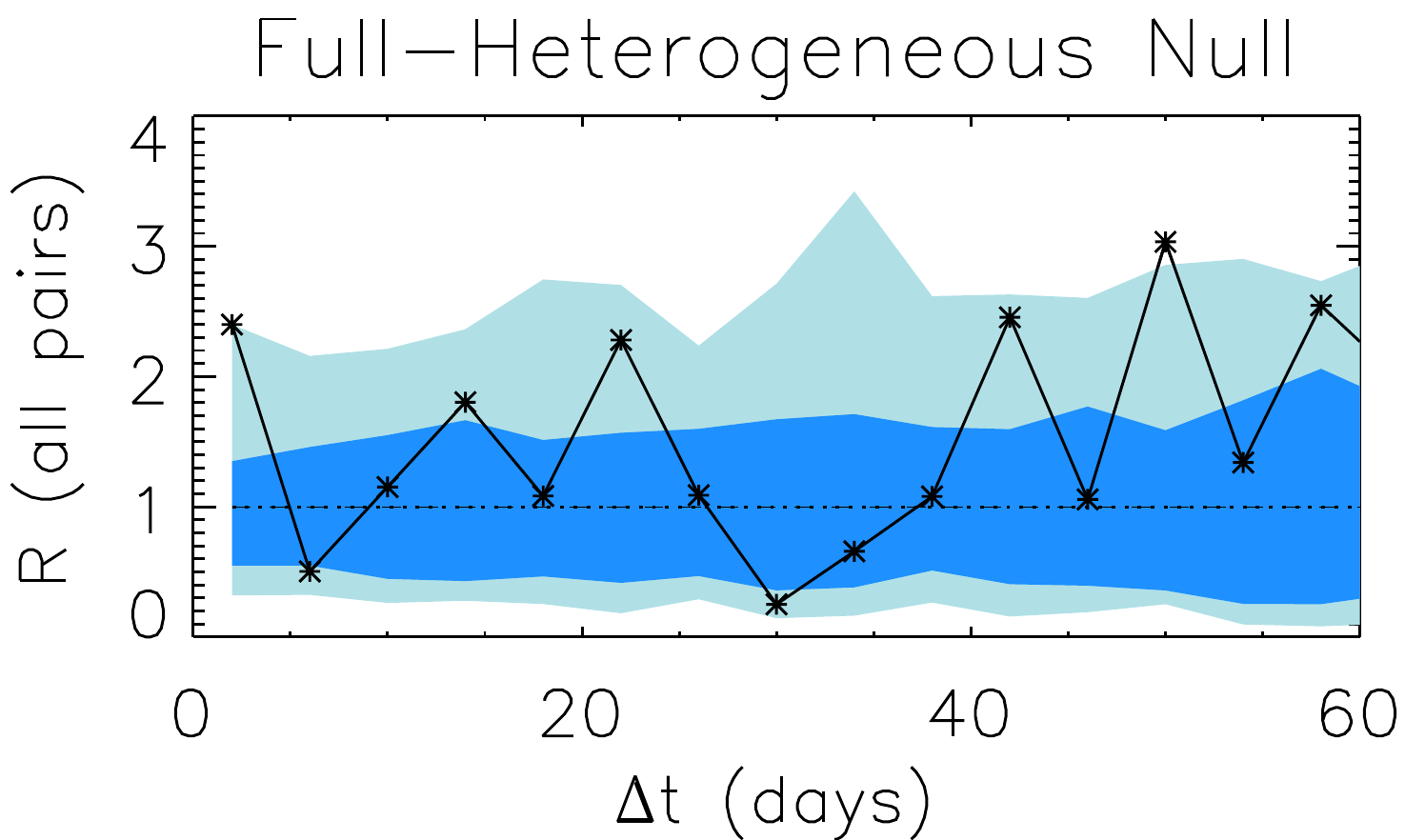}} \\
\multicolumn{2}{c}{ ({\bf c})} \\
\end{tabular}
\caption{Average reciprocity, as a function of separation, $\Delta t$, compared to our null hierarchy. Blue bands indicate the one and two-sigma ranges for the three null models in turn. ({\bf a}): Homogeneous null;  ({\bf b}) Kin-heterogeneous null; ({\bf c}): Full-heterogeneous null. See Section~\ref{hier}for further details on the formulation of these models.}
\label{set-results}
\end{figure}

We present our results for conditional-action reciprocity in Figure~\ref{set-results}. These panels show the average reciprocity, $R$, over all pairs in the system with hosting in both directions ({\em i.e.}, $R(\Delta t)$ is the average of all $R_{ij}(\Delta t)$, such that $N(\textrm{``$i$ hosts $j$''})$ and $N(\textrm{``$j$ hosts $i$''})$ are non-zero and observable at some separation).

Some apparent signatures of conditional-reciprocity can be seen, particularly on the very shortest ($\leq$3 day) scales. Evaluated against the null expectations of no heterogeneity in interaction rates between specific host-guest pairs (Figure~\ref{set-results}a) or of heterogeneity conditioned only on the degree of relatedness (Figure~\ref{set-results}b), the effect is large, with a roughly eight-fold increase in the conditional probability that $j$ hosts $i$ within three days after $i$ hosts $j$. Figure~\ref{set-results}c shows the signal of time-dependent reciprocity compared against a null allowing for idiosyncratic stationary ({\em i.e.}, i.i.d. process) preferences for interaction between host-dyad pairs. Under this stricter null, what appeared to be a large effect reduces in size, but does not disappear entirely: the fully heterogeneous preference model estimates an approximately 2.4-fold increase in the conditional probability that $j$ reciprocates within three days of $i$ hosting $j$, an effect which is significant at $p\approx0.01$, or approximately $2\sigma$.

\begin{figure}[H]
\centering
\begin{tabular}{cc}
\includegraphics[width=3.4in]{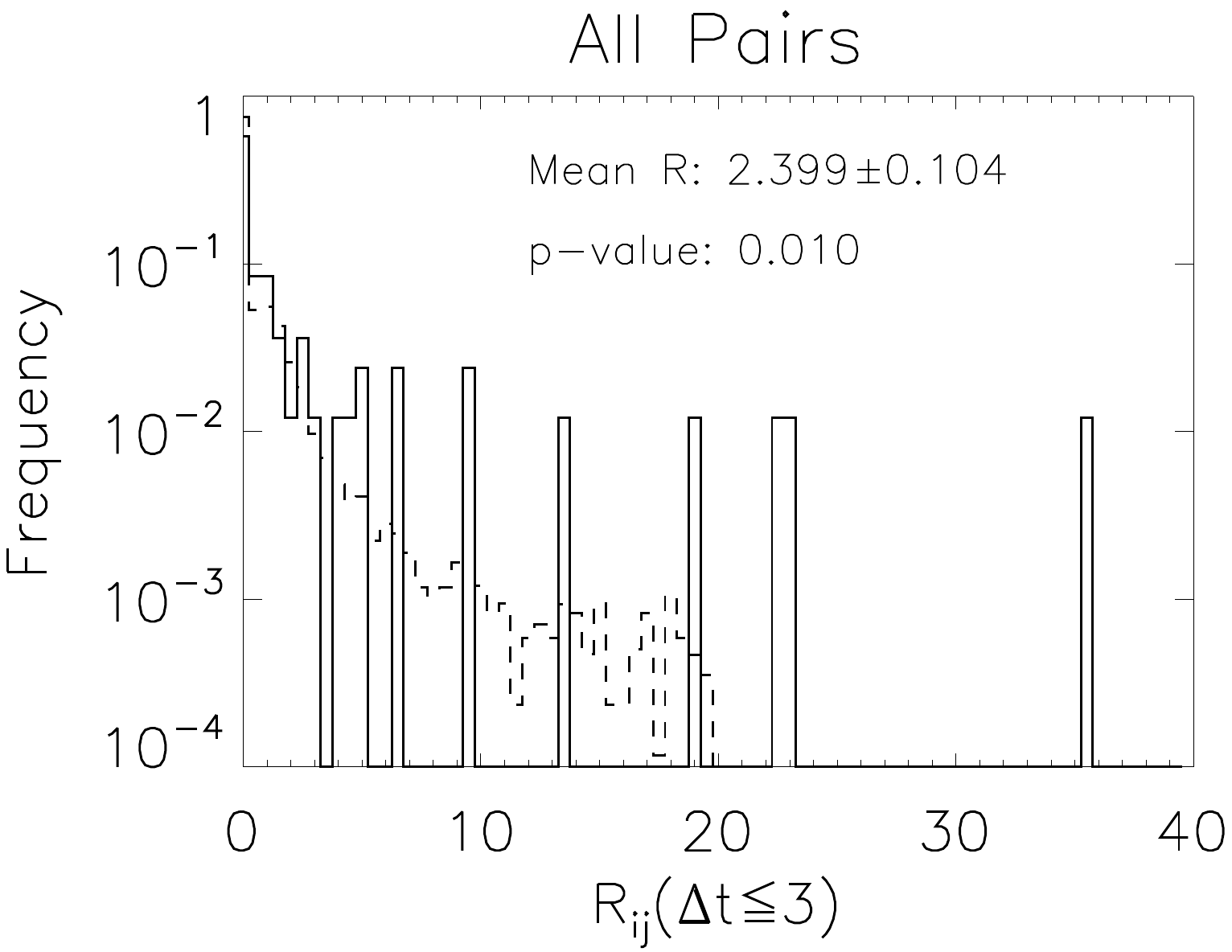} &
\includegraphics[width=3.4in]{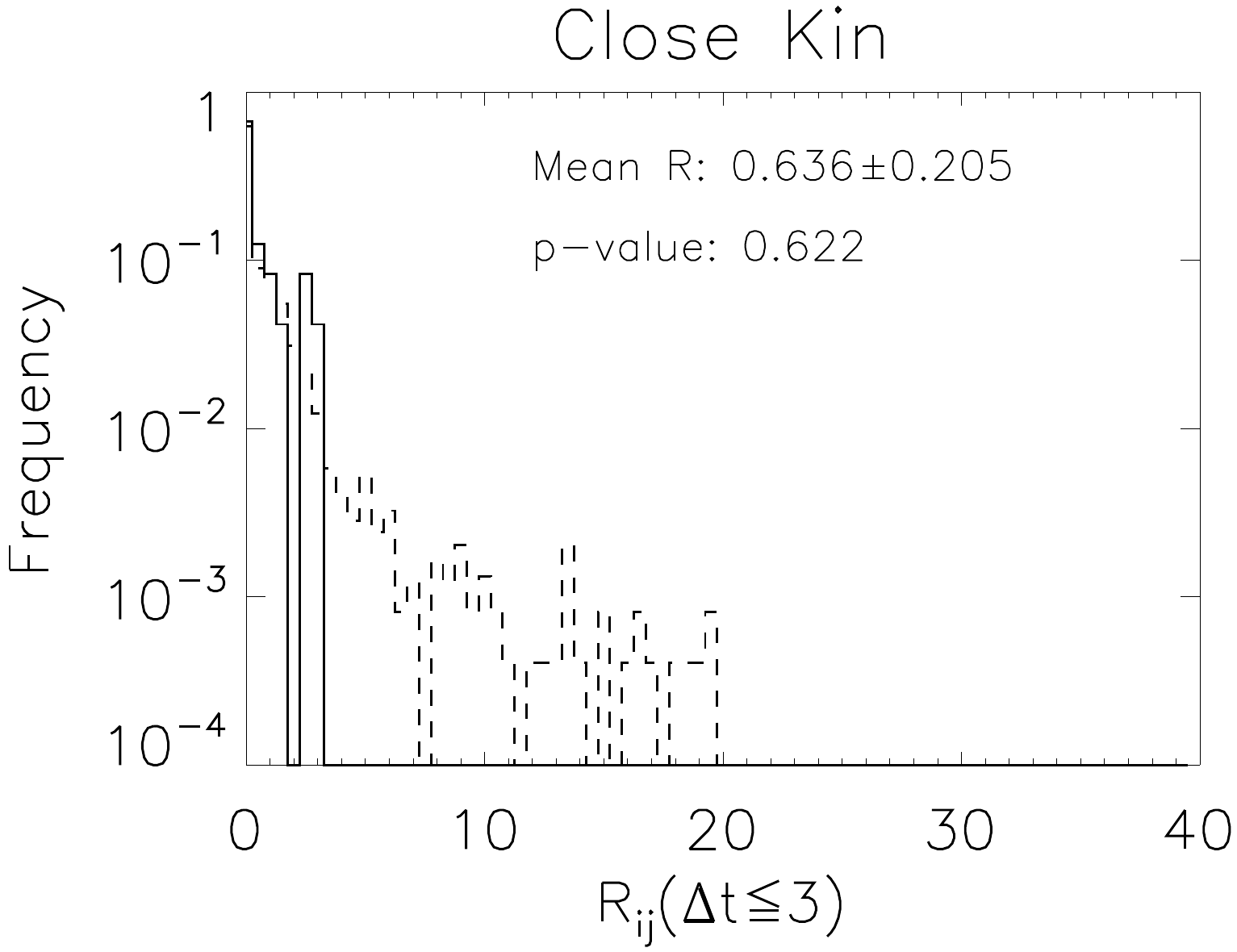} \\
({\bf a}) & ({\bf b}) \\
\multicolumn{2}{c}{
\includegraphics[width=3.4in]{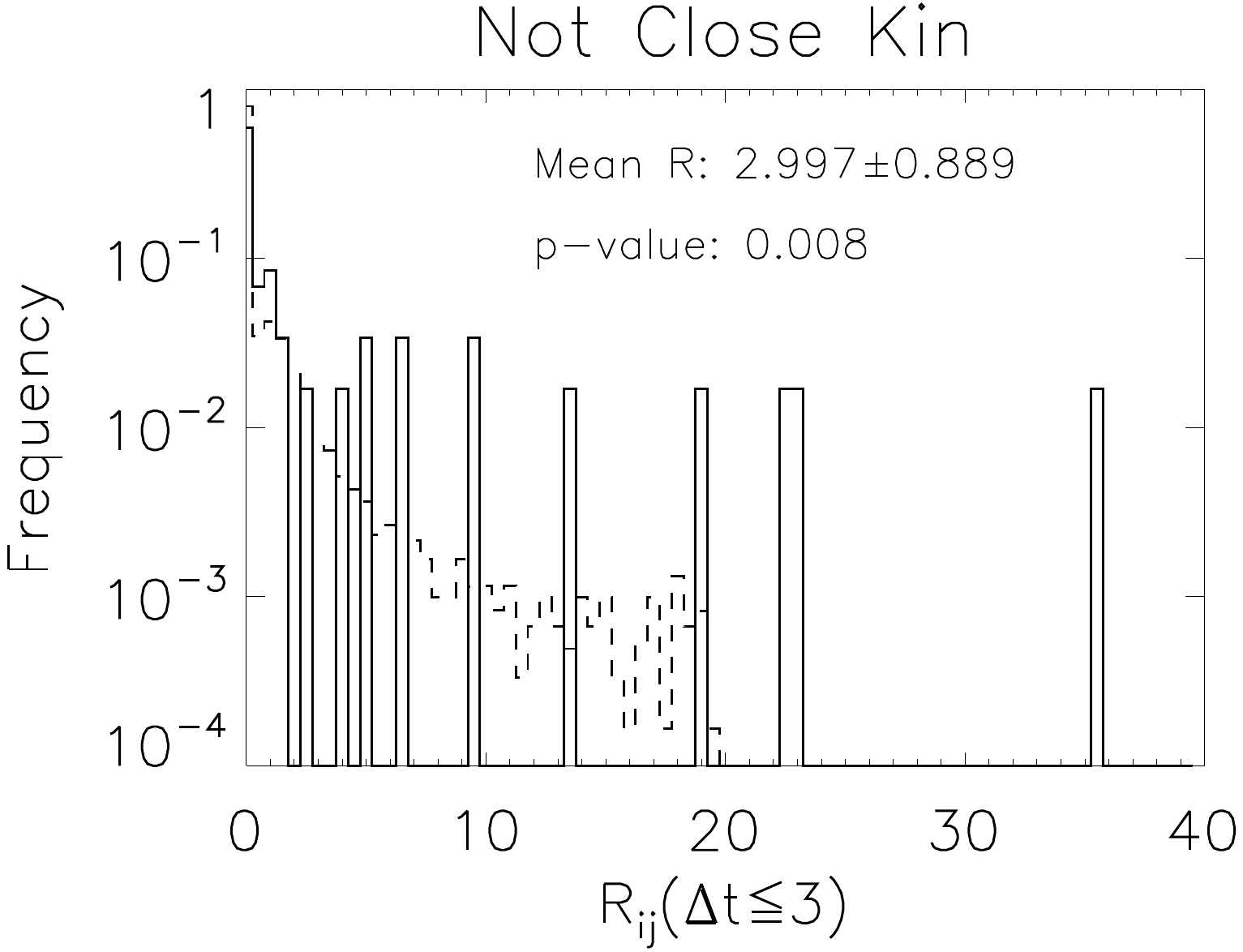}}\\
\multicolumn{2}{c}{ ({\bf c})} \\
\end{tabular}
\caption{Rapid timescale ($t\leq3$~days) reciprocity values. Solid lines show the distribution of observed data; dashed lines that for the full-heterogeneous null. {({\bf a}): All pairs; ({\bf b}): Closely related ($r\ge0.25$) pairs; ({\bf c}): Distant and non-kin ($r<0.25$) pairs.} Close kin show a far weaker signal of reciprocity compared to distant and non-kin.}
\label{set-results2}
\end{figure}

In Figure~\ref{set-results2}, we examine the evidence for short-term ($\leq$3 day) reciprocity among three host-dyad groupings: \emph{All} (all possible hosting pairs), \emph{Close kin} (all pairs with average kinship greater than or equal to 0.25) and \emph{Not close kin} (all pairs with kinship strictly less than 0.25). We plot the distributions for each of these categories, compared to the null expectation, and compute the \textit{p}-score. The positive result for $\leq$3 day conditional-action reciprocity across all pairs (mean $R_{ij}=2.4$, \textit{p}$\approx0.01$) is clearly driven by interactions between distant and non-kin (mean $R_{ij}=$ 3.0, \textit{p} = 0.008) rather than close kin pairs (mean $R_{ij}=0.63$, $p\approx0.6$; no detectable deviation from null expectation). 

The bottom panel of Figure~\ref{set-results} also shows a positive $R$ value exceeding null expectations at the two-sigma level at around 50 days. As there is no specific reason to expect a positive signal of time-dependent reciprocity at this temporal distance (rather than, say, at 45 or 55 days), we consider this most likely a fluctuation due to chance alone. Care must be taken not to hunt for signals at an arbitrary time without either good justification or careful adjustment of \textit{p}-value significance, since, in the words of~\cite{press2007numerical}, ``look long enough, find anything.''

\subsection{Discussion}

We have probed the structure of the Tsimane' manioc beer hosting network for statistical signatures expected from the theories of kin selection and dyadic reciprocity. With respect to kinship, the current results are unambiguous. Kinship independently predicts rates of hosting and attendance between dyads. In addition to the independent effect of kinship, a portion of the variance in hosting attributed to spatial distance can also be attributed to kinship, since kinship also partially determines the location of households.

With respect to reciprocity, the results allow some conclusions, but not others. The stationary analysis of Section~\ref{MLR} shows that across the sample period, families clearly host others that host them in turn at a level above and beyond that expected on the basis of kinship and spatial proximity alone. In the dynamic analysis of Section~\ref{TSA}, we searched for the signature of time-sensitive, conditional-action reciprocity that might account for the observed rates of stationary reciprocation within the network. This analysis revealed a significant increase in the probability that $j$ hosts $i$ within three days of $i$ hosting $j$, above and beyond the general (time-independent) tendency for $j$ to host $i$ over the sample period. The relative size and time-frame of the detected conditional-action reciprocity effect are modest, suggesting that an alternative model---one which allows for dyad-specific preferences for hosting and attendance, but which has no dynamic properties---is in fact \emph{nearly} sufficient to explain the observed rates of bilateral hosting. These results obtain in the context of the four-month baseline of behavior established across the study, a period that included significant changes in task ({e.g.,} a shift from harvesting rice in May and June to clearing new fields in August and September). 

There are a number of reasons why strong signals of conditional-action reciprocity have remained difficult to detect, even if reciprocity does act as an important mechanism maintaining cooperation in this and other social systems. First, it may be that many of the relevant causal events occur outside the confines of the current network. \emph{Chicha}, for example, may be offered as a way of eliciting or rewarding cooperative labor in horticultural work, food sharing or political support. Second, the decision rules underlying conditional social behavior are likely to be substantially more subtle and complex, depending on the specific context and history of interaction, than the simple time-dependent mechanism posited here. Third, the mechanisms supporting reciprocation within long-term relationships---such as selective partner choice and retention or emotional commitment---may occur across relatively longer time scales than those investigated in the current study, creating the conditions for reciprocity without explicit or perceptible score-keeping at shorter time scales \cite{hruschka2010,jaeggi2012}. 

Additionally, the \emph{threat} of conditional-action reciprocity effects may remain relevant to the construction of highly reciprocal and pro-social societies, yet prove undetectable in the absence of perturbation. A perfectly law-abiding society has no record of arrests, but this does not indicate that the laws themselves do not exist. This apparent paradox---that more fully-cooperative societies would show less evidence for dynamical reciprocity---may be accessible to empirical study. While direct-intervention experiments are problematic due to considerations of ethics or expense, naturally occurring perturbations could give a window onto the problem. 

It may be the case, for example, that a new arrival to a highly-cooperative society must establish their \emph{bona fides} with the community by means of conditional-reciprocity effects. In particular, preferences for the new arrival must be established by the other community members, and \textit{vice versa}. In the formal language of the present time-series analysis, a newly arrived family, $i$, could be expected to show evidence for above-null $R$---either $R_{i\star}$ or $R_{\star i}$---as they raise their probability of hosting another family, $j$, in response to signals from $j$ that this is desired. The signal that $j$ might send includes the hosting of $i$ in return, so that first $P(\textrm{``$i$ hosts $j$"})$ rises as $i$ experiments with a new relationship, then $P(\textrm{``$j$ hosts $i$"})$ rises. If the prior probability of $j$ hosting $i$ is small (as might be expected for the new-comer, $i$), this effect would lead to an $R_{ij}$ much greater than unity.

It is worth recognizing that other motivations and processes are likely to underlie behavior in the \emph{chicha} hosting network beyond dyadic reciprocity and kinship-based altruism. Some degree of generosity in this network, for instance, may be motivated by concerns for reputation management, which supersede the specific dyad in question \cite{nowak2005evolution}. Generalized altruism might also be reinforced through mechanisms of reward and punishment not captured here, as suggested by the literature on peer monitoring and strong reciprocity \cite{bowlesgintis2010,bowlesgintis2011}. The fact that one may fairly easily join \emph{chicha} gatherings without explicit invitation may also limit hosts' control of attendance in ways that create a scope for scrounging (or tolerated theft) and disrupt clear patterns of directed reciprocation \cite{gurven2004,blurtonjones1984}. One may note, however, that none of these processes predicts the substantial within-community heterogeneity in dyad-specific relationship quality, which we have observed in the current study. 

The substantial variance in the rates of hosting unexplained by age or family size also begs investigation and is consistent with the hypothesis that hosting may be motivated in part by the benefits of signaling wealth, productivity or social status \cite{smith2000turtle}. Such heterogeneity would also be consistent with a degree of specialization and division of labor between families (with some families focusing, for example, on production of meat, and others on the production of horticultural goods, including \emph{chicha}, as described in \cite{hooper2011,gurvenkaplan2006}).

The current results also bear on the intersection of kin selection and dyadic reciprocity. In addition to the negative (though not significant) interaction between reciprocity and kinship in the stationary analysis, the significant conditional-action reciprocity effect of the dynamic analysis held among distant and non-kin, but not among close kin. These patterns suggest that kinship and reciprocity in this network act as strategic substitutes, rather than complements, in contrast to the results of several evolutionary models and empirical studies \cite{axelrod1981,mcelreath2007,allenarave2008,nolin2011kin}. This may prompt the refinement of the theory to better understand the social and ecological conditions under which different mechanisms supporting altruism and cooperation may reinforce each other or crowd each other out \cite{bowles2012economic}. 

\section{Summary and Conclusion}

In this paper, we have presented new results on the nature of cooperation and the maintenance of social ties and have introduced new methods to detect signals of dynamical processes within social networks. Using detailed information gathered from a sixteen-week ethnographic study of manioc beer drinking events, we have looked for signatures of genetic kinship and stationary reciprocity in patterns of directed social behavior. While spatial distance is a major predictor of the rates of hosting and attendance, families host other families (a) that are more closely related and (b) that host them in turn significantly more frequently than expected based on distance or family-level characteristics alone. 

We have sought a potential dynamical signature for the emergence of these long-lasting stationary ties. Using the formalism of probability calculus, we have searched for the simplest possible signature of \mbox{conditional-action} reciprocity at different time-scales. We have shown how to estimate \mbox{conditional-action} reciprocity and determine its statistical significance using a hierarchy of null behavioral models. The results show a significant signal of \mbox{conditional-action} reciprocity at very short ($\leq$3 day) time scales among distantly and unrelated families, but not among close kin. Apart from this effect, much of the variation in interaction rates is well accounted for by assuming stable preferences for hosting between specific family pairs, which are stationary and unchanging across the time period of observation. 


\section*{\noindent Acknowledgments}
\vspace{12pt}

We are grateful to the community members who participated in this study and to Miguel Mayer and Daniel Vie for their help in collecting data. Thanks to Matthew Schwartz, Adrian Jaeggi, Monique Borgerhoff Mulder, Aaron Blackwell, Bret Beheim, Eric Schniter and three reviewers for their help and insight. This work was supported by {National Science Foundation (NSF)} 
 grant BCS-0422690 and {National Institutes of Health/National Institute on Aging} 
grants R01AG024119-01 and 2P01AG022500-06A1; S.D. was supported by NSF grant EF-1137929, and the Emergent Institutions~Project. 


\section*{\noindent Conflicts of Interest}
\vspace{12pt}

The authors declare no conflict of interest.

\appendix
\section{Appendix: Further Details on Statistical Estimation of Conditional-Action Reciprocity}
\label{app}
\vspace{-12pt}

\subsection{Correction for Seasonality}

Figure~\ref{season-data} shows that the overall level of hosting changes throughout the dataset. While these shifts could in principle be caused by multiple conditional-reciprocity effects, a more parsimonious explanation might associate them with general shifts in activity over the course of the year. A truly i.i.d. sampling from stationary preferences would not allow for these exogenous events and would lead to the anomalous detection of signals of conditional reciprocity. 

As an example, consider the lull in party activity seen in the data in late June. This occurs between a high-activity period in May and a medium-activity period in July and August. If these seasonal effects were not corrected for, this large-scale pattern would lead to a detection of anomalously low conditional reciprocity at separations of roughly thirty days. We thus, in all three nulls, allow for overall, \mbox{family-independent}, shifts in the stationary probabilities in the data. We smooth these corrections on one-week scales, which we consider sufficient to capture exogenously-driven changes in party hosting.

\subsection{Correction for Reporting}

There are two ways in which an observer can come to know about a party: by interviewing the host and asking them to list the guests or by interviewing someone who turns out to have been a guest. The two questions (``who drank at your house?'' and ``at whose house did you drink?''), though formally symmetric (if $i$ hosted $j$, then $j$ was hosted by $i$) may be answered differently, due to memory, shyness or other social factors. In many cases, both host and guest have been interviewed, and differences in the data can \mbox{be reconciled}. 

In other cases, however, only one of the pairs is contacted at that time period, and so, we are unable to perform reconciliation. In order to properly compare simulated data to actual data, however, we need to simulate not only ``the truth''---{\em i.e.}, the actual hosting patterns---but also the reporting of that truth. Since we do not have directly-observed evidence for party attendance and our data rely upon participant report, we model the host reports as truth and deviations from host reports as due to additional \mbox{reporting mechanisms}. 

To a first approximation, we model the guest reporting mechanism with a single parameter, $p_\textrm{omit}$: the probability that a guest fails to report a party that the host reports. We emphasize that this refers not to the forgetting of an actual event (as we do not have access to this information), but rather to the probability that a guest omits describing an event that a host describes.

We can determine $p_\textrm{omit}$ by requiring that our null models reproduce, on average, the same level of realized guest-host pairs (502) as seen in the data. We find that the correction necessary is quite high, $p_\textrm{omit}$, the probability that a guest ``forgets'' to mention a party that the host indicates has taken place, to be $0.687\pm0.001$ for the full-heterogeneous case, $0.720\pm0.002$ for the kin-heterogeneous case and $0.727\pm0.001$ for the homogeneous case.

In our model, this probability is independent of both guest and guest-host. Indirect evidence that this is only an approximation to the true underlying dynamics is given by the slight differences, over and above statistical error, in $p_\textrm{omit}$ within the null hierarchy, which should coincide if (1) the reporting mechanisms are independent of the party-attending mechanisms and (2) all relevant features of the reporting mechanisms are captured.

An additional effect is the possibility of non-random sampling: in particular, a correlation between $i$ hosting $j$, $j$ hosting $i$, $i$ being observed and $j$ being observed. Simulations suggest that this effect, \mbox{if it exists}, is below the level of noise due to small-number fluctuations. In particular, it is not possible to reject the hypothesis that ``correlations do not exist'' by reference to the stationary probabilities observed in the data \textit{versus} those measured in the nulls.
%
\newpage


\begin{thebibliography}{----}
\providecommand{\natexlab}[1]{#1}

\bibitem[Iamele(2001)]{iamele}
Iamele, G.
\newblock {\em Palabras Antiguas y Nuevas del R{\'\i}o Quiquibey en la Amazonia Boliviana} (in Spanish); Programa Regional de Apoyo a los Pueblos Ind{\'\i}ginas de la Cuenca del Amazonas: La Paz, Bolivia, 2001.

\bibitem[Silk(2012)]{silk2012}
Silk, J.B.
\newblock The Adaptive Value of Sociality. In {\em The Evolution of Primate
 Societies}; Mitani, J.C., \mbox{Call, J.}, Kappeler, P.M., Palombit, R.A., Silk,
 J.B., Eds.; University of Chicago Press: {Chicago, IL,  USA,} 2012; pp. 554--564.

\bibitem[Jackson and Watts(2002)]{jackson2002}
Jackson, M.O.; Watts, A.
\newblock The evolution of social and economic networks.
\newblock {\em J. Econ. Theory} {\bf 2002}, {\em 106},~265--295.

\bibitem[Kappeler \em{et~al.}(2013)Kappeler, Barrett, Blumstein, and
 Clutton-Brock]{kappeler2013}
Kappeler, P.M.; Barrett, L.; Blumstein, D.T.; Clutton-Brock, T.H.
\newblock Constraints and flexibility in mammalian social behaviour:
 Introduction and synthesis.
\newblock {\em Philos. Trans. R. Soc. B} {\bf 2013},
 {\em 368}, {20120337}.

\bibitem[Doreian and Stokman(1997)]{doreian1997}
Doreian, P.; Stokman, F.
\newblock {\em Evolution of Social Networks}; Routledge: {Amsterdam, The Netherlands,} 1997; Volume~1.

\bibitem[Newman(2009)]{newman2009}
Newman, M.
\newblock {\em Networks: An Introduction}; Oxford University Press: {Oxford, UK,} 2009.

\bibitem[Flack(2012)]{flack2012}
Flack, J.C.
\newblock Multiple time-scales and the developmental dynamics of social
 systems.
\newblock {\em Philos. Trans. R. Soc. B} {\bf 2012},
 {\em 367},~1802--1810.

\bibitem[DeDeo \em{et~al.}(2010)DeDeo, Krakauer, and Flack]{dedeo1}
DeDeo, S.; Krakauer, D.C.; Flack, J.C.
\newblock {Inductive game theory and the dynamics of animal conflict}.
\newblock {\em {PLoS Comput. Biol.}} {\bf 2010}, {\em 6},~e1000782.

\bibitem[{DeDeo} \em{et~al.}(2011){DeDeo}, {Krakauer}, and {Flack}]{dedeo2}
{DeDeo}, S.; {Krakauer}, D.C.; {Flack}, J.C.
\newblock {Evidence of strategic periodicities in collective conflict
 dynamics}.
\newblock {\em J. R. Soc. Interface} {\bf 2011}, {\em
 8},~1260--1273.

\bibitem[Mowat(1989)]{mowat}
Mowat, L.
\newblock {\em Cassava and Chicha: Bread and Beer of the Amazonian Indians};
 Shire Publications: Aylesbury, UK, 1989.

\bibitem[Hamilton(1964)]{hamilton1964}
Hamilton, W.D.
\newblock The genetical evolution of social behaviour. I.
\newblock {\em J. Theor. Biol.} {\bf 1964}, {\em 7},~1--16.

\bibitem[Trivers(1971)]{trivers1971}
Trivers, R.L.
\newblock The evolution of reciprocal altruism.
\newblock {\em Q. Rev. Biol.} {\bf 1971}, {\textit{46}}, 35--57.

\bibitem[Gurven(2004)]{gurven2004}
Gurven, M.
\newblock To give and to give not: The behavioral ecology of human food
 transfers.
\newblock {\em Behav. Brain Sci.} {\bf 2004}, {\em 27},~543--559.

\bibitem[West \em{et~al.}(2007)West, Griffin, and Gardner]{west2007}
West, S.A.; Griffin, A.S.; Gardner, A.
\newblock Evolutionary explanations for cooperation.
\newblock {\em Curr. Biol.} {\bf 2007}, {\em 17},~R661--R672.

\bibitem[Langergraber(2012)]{langergraber2012}
Langergraber, K.E.
\newblock Cooperation among Kin. In {\em The Evolution of Primate Societies};
 \mbox{Mitani, J.C.,} Call, J., Kappeler, P.M., Palombit, R.A., Silk, J.B., Eds.;
 University of Chicago Press: {Chicago, IL, USA,} 2012; pp. 491--513.

\bibitem[Griffin and West(2003)]{griffin2003}
Griffin, A.S.; West, S.A.
\newblock Kin discrimination and the benefit of helping in cooperatively
 breeding vertebrates.
\newblock {\em Science} {\bf 2003}, {\em 302},~634--636.

\bibitem[Bourke(2011)]{bourke2011}
Bourke, A.F.
\newblock The validity and value of inclusive fitness theory.
\newblock {\em Proc. R. Soc. B} {\bf 2011}, {\em
 278}, 3313--3320.

\bibitem[Service(1962)]{service1962}
Service, E.R.
\newblock {\em Primitive Social Organization}; Random House: {New York, NY, USA,} 1962.

\bibitem[Hames(1987)]{hames1987}
Hames, R.
\newblock Garden labor exchange among the Ye'kwana.
\newblock {\em Ethol. Sociobiol.} {\bf 1987}, {\em 8},~259--284.

\bibitem[Hooper(2011)]{hooper2011}
Hooper, P.L.
\newblock {\em The Structure of Energy Production and Redistribution Among
 Tsimane' \mbox{Forager-Horticulturalists}}; University of New Mexico: {Albuquerque, NM, USA,} 2011.

\bibitem[Koster(2011)]{koster2011interhousehold}
Koster, J.
\newblock Interhousehold meat sharing among Mayangna and Miskito
 horticulturalists in Nicaragua.
\newblock {\em Hum. Nat.} {\bf 2011}, {\em 22},~394--415.

\bibitem[Mauss(1990)]{mauss}
Mauss, M.
\newblock {\em The Gift: The Form and Reason for Exchange in Archaic
 Societies}, W.W. Norton: {New York, NY, USA,} 1990.

\bibitem[Axelrod and Hamilton(1981)]{axelrod1981}
Axelrod, R.; Hamilton, W.D.
\newblock The evolution of cooperation.
\newblock {\em Science} {\bf 1981}, {\em 211},~1390--1396.

\bibitem[Gurven(2006)]{gurven2006}
Gurven, M.
\newblock The evolution of contingent cooperation.
\newblock {\em Curr. Anthropol.} {\bf 2006}, {\em 47},~185--192.

\bibitem[rec()]{recip-note}
Reciprocity might also simply refer to the frequency with which the social
 partners nominated by an individual also nominate that individual in turn;
 here we are interested in reciprocity in networks in which directed edges
 represent tangible flows of benefits, goods, or services between nodes. 

\bibitem[Rao and Bandyopadhyay(1987)]{rao1987measures}
Rao, A.R.; Bandyopadhyay, S.
\newblock Measures of reciprocity in a social network.
\newblock {\em Sankhy{\=a}: Indian J. Stat. Ser. A} {\bf
 1987}, {\textit{49}}, 141--188.

\bibitem[Wasserman and Faust(1994)]{wasserman1994}
Wasserman, S.; Faust, K.
\newblock {\em Social Network Analysis: Methods and Applications}; Cambridge University Press: {Cambridge, UK,} 1994.

\bibitem[Jaeggi and Gurven(2013)]{jaeggireview}
Jaeggi, A.V.; Gurven, M.
\newblock Reciprocity explains food sharing in humans and other primates
 independent of kin selection and tolerated scrounging: A phylogenetic
 meta-analysis.
\newblock {\em Proc. R. Soc. B} {\bf 2013}, {\em 280}, {20131615}.

\bibitem[Gilby(2012)]{gilby2012}
Gilby, I.C.
\newblock Cooperation among Non-Kin: Reciprocity, Markets, and Mutualism. In {\em The Evolution of Primate Societies}; Mitani, J.C., Call, J., Kappeler,
 P.M., Palombit, R.A., Silk, J.B., Eds.; University of Chicago Press: {Chicago, IL, USA,} 2012;
 pp. 514--530.

\bibitem[Jana \em{et~al.}(2013)Jana, Bandyopadhyay, and
 Choudhuri]{jana2013reciprocity}
Jana, R.; Bandyopadhyay, S.; Choudhuri, A.K.
\newblock Reciprocity among farmers in farming system research: Application of
 social network analysis.
\newblock {\em J. Hum. Ecol.} {\bf 2013}, {\em 41},~45--51.

\bibitem[Silk(2003)]{silk2003}
Silk, J.B.
\newblock Cooperation without Counting: The Puzzle of Friendship. In {\em
 Genetic and Cultural Evolution of Cooperation}; Hammerstein, P., Ed.; MIT: {Cambridge, MA, USA,} 2003; pp. 37--54.

\bibitem[Hruschka(2010)]{hruschka2010}
Hruschka, D.J.
\newblock {\em Friendship: Development, Ecology, and Evolution of a
 Relationship}; University of California Press: {Berkeley, CA, USA,} 2010.

\bibitem[Kasper \em{et~al.}(n.d.)Kasper, Fitzherbert, and
 Borgerhoff~Mulder]{kasper}
Kasper, C.; Fitzherbert, E.; Borgerhoff~Mulder, M.
\newblock Who helps and why? Cooperative networks in Mpimbwe.
\newblock {\em Working Paper, University of California Davis} {\bf 2013}.

\bibitem[Hames and McCabe(2007)]{hames2007}
Hames, R.; McCabe, C.
\newblock Meal sharing among the Ye'kwana.
\newblock {\em Hum. Nat.} {\bf 2007}, {\em 18},~1--21.

\bibitem[Hemelrijk(1994)]{hemelrijk1994}
Hemelrijk, C.K.
\newblock Support for being groomed in long-tailed macaques, Macaca
 fascicularis.
\newblock {\em Anim. Behav.} {\bf 1994}, {\em 48},~479--481.

\bibitem[Melis \em{et~al.}(2008)Melis, Hare, and Tomasello]{melis2008}
Melis, A.P.; Hare, B.; Tomasello, M.
\newblock Do chimpanzees reciprocate received favours?
\newblock {\em Anim. Behav.} {\bf 2008}, {\em 76},~951--962.

\bibitem[Koyama \em{et~al.}(2006)Koyama, Caws, and Aureli]{koyama2006}
Koyama, N.; Caws, C.; Aureli, F.
\newblock Interchange of grooming and agonistic support in chimpanzees.
\newblock {\em Int. J. Primatol.} {\bf 2006}, {\em
 27},~1293--1309.

\bibitem[Jaeggi \em{et~al.}(2012)Jaeggi, De~Groot, Stevens, and
 Van~Schaik]{jaeggi2012}
Jaeggi, A.V.; de~Groot, E.; Stevens, J.M.; van~Schaik, C.P.
\newblock Mechanisms of reciprocity in primates: Testing for short-term
 contingency of grooming and food sharing in bonobos and chimpanzees.
\newblock {\em Evol. Hum. Behav.} {\bf 2012}, {\textit{34}, 69--77}.

\bibitem[Schino and Aureli(2009)]{schino2009}
Schino, G.; Aureli, F.
\newblock Reciprocal altruism in primates: Partner choice, cognition, and
 emotions.
\newblock {\em Adv. Stud. Behav.} {\bf 2009}, {\em 39},~45--69.

\bibitem[Frank and Silk(2009)]{frank2009}
Frank, R.E.; Silk, J.B.
\newblock Impatient traders or contingent reciprocators? Evidence for the
 extended time-course of grooming exchanges in baboons.
\newblock {\em Behaviour} {\bf 2009}, {\em 146},~1123--1135.

\bibitem[Gomes \em{et~al.}(2009)Gomes, Mundry, and Boesch]{gomes2009}
Gomes, C.M.; Mundry, R.; Boesch, C.
\newblock Long-term reciprocation of grooming in wild West African chimpanzees.
\newblock {\em Proc. R. Soc. B} {\bf 2009}, {\em
 276},~699--706.

\bibitem[Sabbatini \em{et~al.}(2012)Sabbatini, Vizioli, Visalberghi, and
 Schino]{Sabbatini2012}
Sabbatini, G.; Vizioli, A.D.B.; Visalberghi, E.; Schino, G.
\newblock Food transfers in capuchin monkeys: An experiment on partner choice.
\newblock {\em Biol. Lett.} {\bf 2012}, {\em 8},~757--759.

\bibitem[Tiddi \em{et~al.}(2011)Tiddi, Aureli, di~Sorrentino, Janson, and
 Schino]{Tiddi2011}
Tiddi, B.; Aureli, F.; di~Sorrentino, E.P.; Janson, C.H.; Schino, G.
\newblock Grooming for tolerance? Two mechanisms of exchange in wild tufted
 capuchin monkeys.
\newblock {\em Behav. Ecol.} {\bf 2011}, {\textit{22},} 663--669.

\bibitem[McElreath and Boyd(2007)]{mcelreath2007}
McElreath, R.; Boyd, R.
\newblock {\em Mathematical Models of Social Evolution: A Guide for the
 Perplexed}; University of Chicago Press: {Chicago, IL, USA,} 2007.

\bibitem[McGlothlin \em{et~al.}(2010)McGlothlin, Moore, Wolf, and
 Brodie~III]{mcglothlin2010interacting}
McGlothlin, J.W.; Moore, A.J.; Wolf, J.B.; Brodie, E.D., III.
\newblock Interacting phenotypes and the evolutionary process. III. Social
 evolution.
\newblock {\em Evolution} {\bf 2010}, {\em 64},~2558--2574.

\bibitem[Ak{\c{c}}ay and Van~Cleve(2012)]{akccay2012behavioral}
Ak{\c{c}}ay, E.; van~Cleve, J.
\newblock Behavioral responses in structured populations pave the way to group
 optimality.
\newblock {\em Am. Nat.} {\bf 2012}, {\em 179},~257--269.

\bibitem[Allen-Arave \em{et~al.}(2008)Allen-Arave, Gurven, and
 Hill]{allenarave2008}
Allen-Arave, W.; Gurven, M.; Hill, K.
\newblock Reciprocal altruism, rather than kin selection, maintains nepotistic
 food transfers on an Ache reservation.
\newblock {\em Evol. Hum. Behav.} {\bf 2008}, {\em 29},~305--318.

\bibitem[Nolin(2011)]{nolin2011kin}
Nolin, D.A.
\newblock Kin preference and partner choice.
\newblock {\em Hum. Nat.} {\bf 2011}, {\em 22},~156--176.

\bibitem[Nowak and Sigmund(2005)]{nowak2005evolution}
Nowak, M.A.; Sigmund, K.
\newblock Evolution of indirect reciprocity.
\newblock {\em Nature} {\bf 2005}, {\em 437},~1291--1298.

\bibitem[Macfarlan \em{et~al.}(2012)Macfarlan, Remiker, and
 Quinlan]{macfarlan2012competitive}
Macfarlan, S.J.; Remiker, M.; Quinlan, R.
\newblock Competitive altruism explains labor exchange variation in a Dominican
 community.
\newblock {\em Curr. Anthropol.} {\bf 2012}, {\em 53},~118--124.

\bibitem[Gurven \em{et~al.}(2000)Gurven, Allen-Arave, Hill, and
 Hurtado]{gurven2000wonderful}
Gurven, M.; Allen-Arave, W.; Hill, K.; Hurtado, M.
\newblock “It's a wonderful life”: Signaling generosity among the Ache of
 Paraguay.
\newblock {\em Evol. Hum. Behav.} {\bf 2000}, {\em 21},~263--282.

\bibitem[Bowles and Gintis(2010)]{bowlesgintis2010}
Bowles, S.; Gintis, H.
\newblock The evolution of strong reciprocity: Cooperation in heterogeneous
 populations.
\newblock {\em Theor. Popul. Biol.} {\bf 2010}, {\em 65},~17--28.

\bibitem[Bowles and Gintis(2011)]{bowlesgintis2011}
Bowles, S.; Gintis, H.
\newblock {\em A Cooperative Species: Human Reciprocity and its Evolution};
 Princeton University Press: {Princeton, NJ, USA,} 2011.

\bibitem[Smith and Bird(2000)]{smith2000turtle}
Smith, E.A.; Bird, R.L.B.
\newblock Turtle hunting and tombstone opening: Public generosity as costly
 signaling.
\newblock {\em Evol. Hum. Behav.} {\bf 2000}, {\em 21},~245--261.

\bibitem[Gintis \em{et~al.}(2001)Gintis, Smith, and Bowles]{gintis2001costly}
Gintis, H.; Smith, E.A.; Bowles, S.
\newblock Costly signaling and cooperation.
\newblock {\em J. Theor. Biol.} {\bf 2001}, {\em 213}, 103--119.

\bibitem[Gurven \em{et~al.}(2012)Gurven, Stieglitz, Hooper, Gomes, and
 Kaplan]{gurven2012womb}
Gurven, M.; Stieglitz, J.; Hooper, P.L.; Gomes, C.; Kaplan, H.
\newblock From the womb to the tomb: The role of transfers in shaping the
 evolved human life history.
\newblock {\em Exp. Gerontol.} {\bf 2012}, {\em 47},~807--813.

\bibitem[Gurven \em{et~al.}(2007)Gurven, Kaplan, and Zelada Supa]{gurven2007mortality}
Gurven, M.; Kaplan, H.; Zelada Supa, A.
\newblock Mortality experience of Tsimane Amerindians of Bolivia: Regional
 variation and temporal trends.
\newblock {\em Am. J. Hum. Biol.} {\bf 2007}, {\em
 19},~376--398.

\bibitem[Therneau \em{et~al.}(2013)Therneau, Atkinson, Sinnwell, Matsumoto,
 Schaid, and McDonnell]{kinship2}
Therneau, T.; Atkinson, E.; Sinnwell, J.; Matsumoto, M.; Schaid, D.; McDonnell, S.
\newblock {\em kinship2: Pedigree Functions}, R Package Version 1.5.0.; 
\newblock 2013.
Available online: http://CRAN.R-project.org/package=kinship2 (accessed on 1 July 2013). 

\bibitem[{R Core Team}(2013)]{rlang}
{The R Core Team}.
\newblock {\em R: A Language and Environment for Statistical Computing}, version 3.0.1;
\newblock R Foundation for Statistical Computing: Vienna, Austria, 2013. 
Available online: http://www.R-project.org/ (accessed on 1 September 2013).

\bibitem[Csardi and Nepusz(2013)]{igraph}
Cs\'{a}rdi, G.; Nepusz, T.
\newblock {\em igraph: The igraph Software Package for Complex Network
 Research}, \linebreak R Package version 0.6.5; 2013;
\newblock Available online: http://igraph.sf.net (accessed 1 September 2013).

\bibitem[Bates \em{et~al.}(2013)Bates, Maechler, and Bolker]{lme4}
Bates, D.; Maechler, M.; Bolker, B.
\newblock {\em lme4: Linear Mixed-Effects Models Using S4 Classes}, R~Package version 0.999999-2; 2013.
\newblock Available online: http://CRAN.R-project.org/package=lme4 (accessed 1 September 2013).

\bibitem[mod()]{model-note}
If it were the case that family $i$ hosting family $j$ on a particular day made
 it more difficult to host another family $k$ on the same day ({e.g.,} due
 to limited space or \emph{chicha}), then an alternative specification of
 Equation~\eqref{logistic}, such as the multinomial logistic, might be preferred in
 order to represent the negative interdependence of the two dyadic
 interactions. However, since two or more families (up to a maximum of 10)
 were hosted on 106 out of 193 risk days that one family hosted at least one
 other family, the multinomial specification---which forces an either-or
 choice between families to host---would be inappropriate to apply to the
 current dataset. We thank one of the reviewers for these insights. 

\bibitem[Gelman and Hill(2007)]{gelman2007data}
Gelman, A.; Hill, J.
\newblock {\em Data Analysis Using Regression and Multilevel/Hierarchical
 Models}; Cambridge University Press: {Cambridge, UK,} 2007.

\bibitem[Hox(2010)]{hox2010multilevel}
Hox, J.
\newblock {\em Multilevel Analysis: Techniques and Applications}; Routledge
 Academic: {New York, NY, USA,} 2010.

\bibitem[Pearl(2009)]{pearl}
Pearl, J.
\newblock {\em Causality}, 2nd ed.; Cambridge University Press: {Cambridge, UK,} 2009.


\bibitem[dat()]{data-note}
The frequencies of hosting and attendance are also associated with household
 distance from village center ($\beta=-0.816$, $p=0.239$ for hosts;
 $\beta=-1.003$, $p=0.001$ for guests) and mean relatedness to other families
 in the community ({\em i.e.}, node strength in the kinship network; $\beta=29.58$,
 $p=0.119$ for hosts; $\beta=23.75$, $p<0.001$ for guests); these
 relationships, however, are more directly captured (and statistically
 obviated: $p>0.6$) by the dyad-specfic distance and relatedness terms. 

\bibitem[Press(2007)]{press2007numerical}
Press, W.
\newblock {\em Numerical Recipes: The Art of Scientific Computing}, 3rd ed.;
 Cambridge University Press: {Cambridge, UK,} 2007.

\bibitem[Blurton~Jones(1984)]{blurtonjones1984}
Blurton~Jones, N.G.
\newblock A selfish origin for human food sharing: Tolerated theft.
\newblock {\em Ethol. Sociobiol.} {\bf 1984}, {\em 5},~1--3.

\bibitem[Gurven and Kaplan(2006)]{gurvenkaplan2006}
Gurven, M.; Kaplan, H.
\newblock Determinants of time allocation across the lifespan: A theoretical
 model and an application to the Machiguenga and Piro of Peru.
\newblock {\em Hum. Nat.} {\bf 2006}, {\em 17},~1--49.

\bibitem[Bowles and Polania-Reyes(2012)]{bowles2012economic}
Bowles, S.; Polania-Reyes, S.
\newblock Economic incentives and social preferences: Substitutes or
 complements?
\newblock {\em J. Econ. Lit.} {\bf 2012}, {\em 50},~368--425.

\end{thebibliography}

\end{document}